\documentclass[onecolumn,noshowpacs]{revtex4}

\usepackage{graphicx}%
\usepackage{dcolumn}
\usepackage{amsmath}
\usepackage{bm}
\usepackage{braket}

\makeatletter
\def\btt#1{\texttt{\@backslashchar#1}}%
\DeclareRobustCommand\bblash{\btt{\@backslashchar}}%
\makeatother

\begin{document}


\title{Two-particle asynchronous quantum correlation: wavefunction collapse acting as a beamsplitter}

\author{F.V. Kowalski and R.S. Browne}
\affiliation{Physics Department, Colorado
School of Mines, Golden CO. 80401 U.S.A.}

\begin{abstract}
A two-body quantum correlation is calculated for a particle reflecting from a moving mirror. Correlated interference results when the incident and reflected particle substates and their associated mirror substates overlap. Using the Copenhagen interpretation of measurement, an asynchronous joint probability density (PDF), which is a function both of the different positions and different times at which the particle and mirror are measured, is derived assuming that no interaction occurs between each measurement. Measurement of the particle first, in the correlated interference region, results in a splitting of the mirror substate into ones which have and have not reflected the particle. An analog of the interference from the Doppler effect for only measurements of the particle (a marginal PDF), in this two-body system, is shown to be a consequence of the asynchronous measurement. The simplification obtained for a microscopic particle reflecting from a mesoscopic or macroscopic mirror is used to illustrate asynchronous correlation interferometry. In this case, the small displacement between these mirror states can yield negligible environmental decoherence times. In addition, interference of these mirror states does not vanish in the limit of large mirror mass due to the small momentum exchange in reflecting a microscopic particle.

\end{abstract}

\pacs{03.65.Ta, 03.65.-w,03.75.Dg,03.75.-b}

\maketitle

\section{Introduction}
\label{sec:intro}

Correlation and interference distinguish quantum from classical physics. The former is manifest in the measurement of many-body coincidences predicted by a quantum joint probability density function (PDF). Some observable correlations cannot be realized classically \cite{bell}. Quantum interference is most familiar as a one-body PDF for an outcome that can be achieved in at least two indistinguishable ways. However, it can also be generated by superposing many-body states in indistinguishable ways \cite{Gottfried}. 

A many-body interferometer is more difficult to treat since it is possible to measure correlations in the probability of finding each particle at different positions {\em and} times. It differs from a one-body interferometer primarily in its beamsplitting mechanism. Here, correlation and interference are combined in asynchronous correlation interferometry of a bipartite system, using measurement of one substate as the beamsplitter for the other.

Experimental confirmation of quantum correlation has involved photons \cite{Salart,Gröblacher}, atoms \cite{Wineland}, and Josephson phase qubits \cite{Ansmann}. However, there is little experimental evidence of correlated interference between massive particles.

Analysis of correlated systems usually begins with an expression for the many-body state after it has been prepared by an interaction. For bipartite systems this is routinely done with photon pairs from parametric downconversion. The extremes of either a few or a large number of particles are most often treated. In the former case the analysis typically involves a simultaneous correlation between the positions \cite{Gottfried} or angular momenta of two particles \cite{aspect,bohm}. The formalism often deals only with identical particles. Here a mechanism is describe which generates interferometric correlations between distinguishable particles, which can be of disparate masses.

The issue of non-simultaneous correlations is explored here in the context of perhaps the simplest quantum correlation possible: a particle reflecting elastically from a mirror, both of which are distinguishable. The mirror and particle have non-zero rest mass and motion is in free space along one dimension, with all states unbound. Measurements of particle reflection, but not associated with correlated interference, have involved mirrors that reflect atoms \cite{kouznetsov} and Bose-Einstein condensates \cite{pasquini}, atoms reflecting from a solid surface \cite{shimizu}, neutrons \cite{hils} and atoms \cite{colombe} reflecting from vibrating mirrors, and atoms reflecting from a switchable mirror \cite{szriftgiser}.

Consider first the particle and mirror in uncorrelated eigenstates of energy before reflection, referred to as the `incident harmonic state'. The energy eigenstate after reflection, referred to as the `reflected harmonic state', results in correlation between the particle and mirror via conservation of energy and momentum: an energy or momentum measurement of the reflected particle yields a correlated energy or momentum measurement of the mirror when given the incident harmonic state. Superposing such states yields the incident and reflected particle-mirror wavegroups. These differ from the harmonic states in that a measurement of the energy (or position) of the particle does not uniquely constrain the energy (or position) of the mirror since the reflected particle-mirror wavegroup is not in an eigenstate of the energy (or position) operator. 

Such incident and reflected wavegroup particle-mirror states interfere when they overlap. This is similar to the transient one-body interference of an electromagnetic wavegroup reflecting from a stationary mirror \cite{Wiener}. However, classically the mirror experiences only a continuous force due to radiation pressure.

Quantum mechanically, interference occurs since the incident and reflected states are indistinguishable for a measurement of position (but not for a momentum measurement). Interference is expected between the incident and reflected particle substates {\em along with} interference of the mirror substates which have and have not reflected the particle. Their correlation is perhaps not expected, being a consequence of the solution to the Schr\"odinger equation from which a joint PDF is constructed. The correlations in the two-body interference are manifest as coincidence rates, e.g. a correlation in the {\em simultaneous} measurement of particle and mirror positions.

This two-body wavefunction is then modified to incorporate predictions for measurement of the particle first and then later that of the mirror, using the Copenhagen interpretation of measurement in quantum mechanics. The resulting PDF is a function both of the different particle positions and different times at which each is measured. An assumption used is that between the times of the two measurements, there is neither interaction between the particle and mirror nor with the environment. 

The focus of the discussion is on asynchronous correlated interference. In this case, a measurement of only the particle in the correlated interference region splits the mirror substate into ones which have and have not reflected the particle. Later measurement of the mirror reveals this interference.

\begin{center}
\begin{figure}
\includegraphics[scale=0.29]{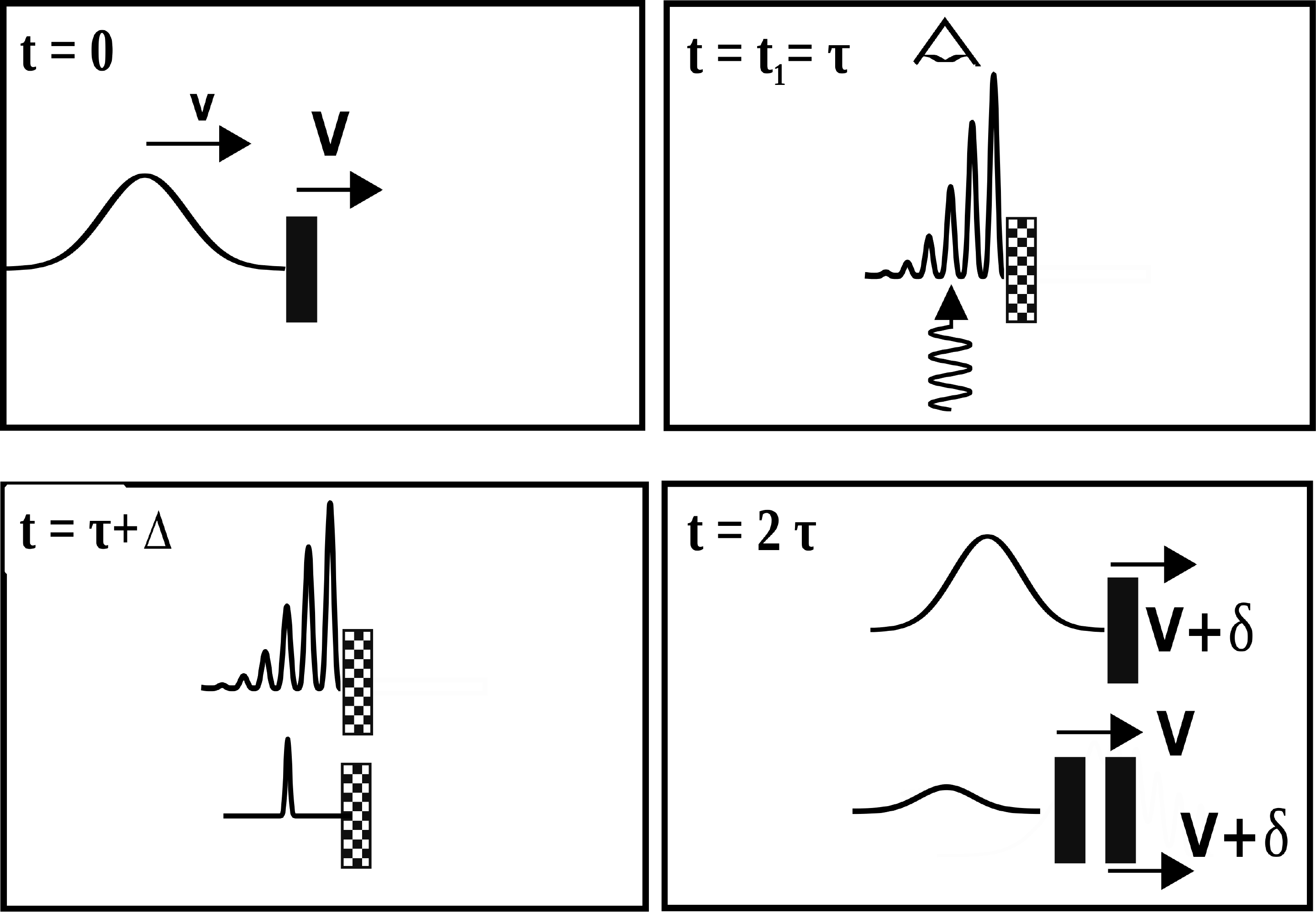}
\caption{Asynchronous measurement schematic. A particle wavegroup reflects from a moving mirror. The `mirror' is represented as the black rectangle, moving to the right. The effects of measuring the particle in the interference region are contrasted with no such measurement in the bottom two schematics.}
\label{fig:overview}
\end{figure}
\end{center}

Fig. \ref{fig:overview} is a schematic representation of this state splitting process. Actual two-body PDFs are shown in later sections. The particle PDF before interaction is the Gaussian at $t=0$, moving to the right at speed $v$, while the mirror is represented as the black rectangle, moving to the right at speed $V<v$. At $t=t_{1}=\tau$ interference between the incident and reflected particle substates is shown as oscillations in its PDF while interference between the mirror substates which have and have not reflected the particle is represented by the rectangular checkerboard pattern rather than the solid rectangle. This then is a region of correlated interference, although in this schematic the correlation cannot be illustrated. For simplicity, speeds are not shown in this and the next snapshot. 

Measurement of the particle but not the mirror is represented by the `photon' coming from below, interacting with the particle, and then being measured by the detector above at time $t_{1}$. More details on this aspect of the process are given in section \ref{sec:measurementtheory}.

The lower two schematics illustrate the states with and without this measurement, using the lower and upper halves of each schematic, respectively. Without measurement, the particle wavegroup reflects from the mirror and continues to move to the right with reduced speed while the speed of the mirror increases. Without measurement, the correlated interference disappears when the wavegroups no longer overlap as illustrated when $t=2\tau$. In addition, only the mirror state which reflects the particle survives (as is illustrated in the upper half of the $t=2\tau$ schematic). This is a consequence of the complete particle wavegroup having reflected.

A noise-free and non-destructive measurement of only the particle, however, collapses the particle substate as shown in the lower portion of the $t=\tau+\Delta$ schematic, with $\Delta<<\tau$. Yet the mirror remains in a superposition state. At time $t=2\tau$ the two mirror states have separated a distance greater than the wavegroup size due to their different speeds and are then represented by solid rectangles. Although there is then no interference, the mirror remains in a superposition of having both reflected and not reflected the particle for all later times. This splitting is similar to a beamsplitter producing photon states which traverse spatially separate and therefore distinguishable paths. Finally, it should be noted that the particle state after measurement spreads, as shown in the lower portion of the $t=2\tau$ schematic. It is assumed that this spreading wavegroup does not interact with the split mirror states for times $t>\tau$.

To determine the consequences of such measurements, a two-body solution of the Schr\"odinger equation for simultaneous measurement is first obtained for the particle-mirror system by applying standard techniques to derive the energy eigenstates. This is used to construct the asynchronous two-body wavefunction. Wavegroups are formed from a superposition of these solutions. Particular emphasis is given to correlated interference or the overlap region of the incident and reflected particle-mirror wavegroups. A marginal PDF is illustrated, using an example of interference for the particle when the mirror is not measured, which yields an analog of the Doppler effect for the interference of light in retro-reflection from a moving mirror. The remainder of the paper deals predominately with issues which illustrate the unique properties of asynchronous correlation interferometry, which are most easily understood using the example of a microscopic particle reflecting from a mesoscopic/macroscopic mirror. This example is not intended as a practical experimental proposal. Rather, observation of asynchronous correlations will more likely first occur with a microscopic particle reflecting from a ``microscopic'' mirror.

\section{Particle reflecting from a mirror}
\label{sec:Theory}

\subsection{Simultaneous measurement}
\label{sec:theoryparticlemirror}

The particle-mirror interaction is modeled as a moving delta function potential where reflection is assumed to occur at the center of mass of the mirror with the Schr\"odinger equation given by
\begin{eqnarray}
(\hbar^{2} \partial_{x_{1}}^{2}/2m+\hbar^{2} \partial_{x_{2}}^{2}/2M+\beta \delta[x_{1}-x_{2}]+i\partial_{t})\Psi=0,
\label{eq:Schreqn}
\end{eqnarray}
where square brackets are used to indicate the argument of a function and $x_{1}$ and $x_{2}$ are the particle and mirror positions along the x-axis. The mirror reflectivity, related to $\beta$, goes to infinity for a lossless mirror. 

The standard separable solution to this equation results from a center of mass (cm) and relative (rel) transformation of the particle-mirror system (not to be confused with the cm of the particle or mirror). This does not change the total energy  $E=(\hbar K)^{2}/(2M)+(\hbar k)^{2}/(2m)=E_{rel}+ E_{cm}$, where $k$ and $K$ are the wavevectors for the particle and mirror with $k=m v/\hbar$, $K=M V/\hbar$, masses $m$, $M$, and initial velocities $v$ and $V$, respectively. The transformed Schr\"odinger equation becomes
\begin{eqnarray}
(\frac{\hbar^{2} \partial_{x_{cm}}^{2}}{2M_{tot}}+\frac{\hbar^{2} \partial_{x_{rel}}^{2}}{2 \mu}+\beta \delta[x_{rel}] \notag
+i\partial_{t})\Psi[x_{cm},x_{rel},t]=0
\label{eq:Scheqncm}
\end{eqnarray}
where $M_{tot}=m+M$, $\mu=mM/(m+M)$, $x_{cm}=(mx_{1}+Mx_{2})/M_{tot}$, and $x_{rel}=x_{1}-x_{2}$. 

The next step in the derivation requires parsing the energy into $E=E_{cm}+E_{rel}$ and assuming the separable solution
\begin{eqnarray}
\Psi[x_{cm},x_{rel},t]=\psi_{cm}[x_{cm},t] \psi_{rel}[x_{rel}, t]  \notag \\ =e^{-i E_{cm} t/\hbar} U[x_{cm}] e^{-i E_{rel} t/\hbar}u[x_{rel}],
\label{eq:Schtot}
\end{eqnarray}
which reduces the Schr\"odinger equation into two ordinary differential equations,
\begin{equation}
-\frac{\hbar^{2}}{2M_{tot}} \frac{d^{2}U[x_{cm}]}{dx_{cm}^{2}} = E_{cm} U[x_{cm}] 
\label{eq:ScheqODE1}
\end{equation}
\begin{equation}
-\frac{\hbar^{2}}{2 \mu} \frac{d^{2}u[x_{rel}]}{dx_{rel}^{2}}+\beta \delta[x_{rel}] = E_{rel} u[x_{rel}].
\label{eq:ScheqODE2}
\end{equation}

The solution to these equations is then obtained from the initial values and boundary conditions. Finally, this solution is transformed back to the particle-mirror system yielding $\Psi[x_{1},x_{2},t]$. 

An example of this procedure is found in the solution to the hydrogen atom where the Schr\"odinger equation is first transformed from the laboratory to the cm-relative coordinates yielding {\em uncorrelated} substates. Transforming back to the electron-proton system yields {\em correlated} electron and proton substates \cite{Tommasini}.

In subsection \ref{sec:non-simultaneous}, the infinite potential energy boundary condition at the mirror surface is used to obtain energy eigenstate solutions to eqns. \ref{eq:ScheqODE1} and \ref{eq:ScheqODE2}. The full solution is then transformed to the particle-mirror system. In subsection \ref{sec:wavegroupsimultaneous}, wavegroups are formed via a superposition of these states.

\subsection{Overview of the asynchronous method}
\label{sec:overview}

This solution, $\Psi[x_{1},x_{2},t]$, is next used to construct the wavefunction which predicts the outcome of asynchronous measurements. For illustrative purposes, let the particle be measured first. Experimental realizations of such a measurement procedure are discussed in subsection \ref{sec:measurementtheory}.

To account for different temporal measurements of the particle and mirror, respective time parameters $t_{1}$ and $t_{2}$ are used instead of $t$ in $\Psi[x_{1},x_{2},t]$. Coefficients of these time parameters in the phase are then the energies of the particle and mirror, respectively. 

This is a parsing of the energy in the solution to the Schr\"odinger equation for the particle-mirror system, similar to that applied after eqn. \ref{eq:Schreqn} to transform it into the cm and rel system. Since that transformation yielded a simultaneous PDF there was no need for the two-time notation. To modify it for asynchronous predictions, relative and cm times are introduced in subsection \ref{sec:non-simultaneous}, which then both account for different temporal measurements of the cm and relative positions and isolate the cm and rel energy terms. 

However, transformation to the cm-rel system is simply a mathematical technique to facilitate a solution to the Schr\"odinger equation. For example, there exists no observable particle with a reduced mass as described in the cm-rel system.  Once a solution, which satisfies the initial conditions and boundary values, is obtained in the cm-rel system the inverse transformation is applied to describe the system that can be measured. 

These time parameters provide labels for the particle and mirror energies in the wavefunction's phase and also label the different times at which the particle and mirror are measured. They assist in isolating the energy of the particle from that of the mirror, just as $x_{1}$ and $x_{2}$ isolate the effect of the particle and mirror wavevectors in the phase of the wavefunction. This then allows all particle parameters to be fixed in the two-body wavefunction when the particle is measured. Such labeling, while simple in a separable system, is not trivial in the correlated particle-mirror system. 

The different time labels do not indicate evolution of the subsystems via different Hamiltonians as do the different time variables used by McGuire \cite{Mcguire} nor are they used to generate a manifestly Lorentz invariant theory as done by Petrat \cite{petrat}. They neither tick at different rates nor are they out of phase, but rather act as only a label, just as $x_{1}$ and $x_{2}$ label the particle and mirror spatial coordinates along the $x$ axis. Such notation then results in the wavefunction derived in subsection \ref{sec:theoryparticlemirror} being expressed as $\Psi[x_{1},x_{2},t] \rightarrow \Psi[x_{1},t_{1},x_{2},t_{2}]$.

Now let the particle be measured first, at position $x_{1}=x_{10}$ at time $t_{10}$. This then collapses the particle substate into an eigenstate of that operator, $\Psi[x_{1},t_{10}]=\delta[x_{1}-x_{10}]$. The first measurement forces the two-body correlated wavefunction into an uncorrelated state which is, at the time that the particle is measured, a product of the particle substate with the mirror substate, given by $\Psi[x_{1},t_{10}] \Psi[x_{10},t_{10},x_{2},t_{10}]$.

The particle state after measurement is irrelevant to the subsequent time evolution of the mirror since they are no longer correlated and no longer interact. The mirror state is determined by fixing all the particle coordinates in the two-body wavefunction at the time of the measurement of the particle. The two time formalism facilitates this procedure by fixing the particle parameters while allowing only those of the mirror to time evolve. The coordinates of this measurement of the particle, $(x_{10},t_{10})$, then determine the initial state of the mirror from the two-body wavefunction as $\Psi[x_{10},t_{10},x_{2},t_{10}]$. This one-body wavefunction for the mirror then continues to time evolve with $\Psi[x_{10},t_{10},x_{2},t_{2}]$ for $t_{2}>t_{10}$.

For example, the phase of the mirror substate evolves in time only with the mirror's energy times the time $t_{2}$. The particle's energy times the time $t_{1}$ term in the phase should not influence the time evolution of the mirror substate after the particle has been measured. It does not in this formalism since all of the energy terms associated with the particle are fixed at $t_{1}=t_{10}$. Similarly, the particle's wavevector times the position $x_{1}$ is fixed by setting $x_{1}=x_{10}$ while the mirror's wavevector times its position $x_{2}$ is allowed to vary.

The probability of {\em simultaneously} measuring the particle and mirror within a small region $\Delta x_{1} \Delta x_{2}$ around $(x_{10},x_{20})$, over which the PDF is essentially constant, is $\textmd{Pr}[x_{10}-\Delta x_{1}<x_{1}<x_{10}+\Delta x_{1},x_{20}-\Delta x_{2}<x_{2}<x_{20}+\Delta x_{2}]=\textmd{PDF}\Delta x_{1} \Delta x_{2}$ with $\textmd{PDF}=\Psi[x_{10},x_{20},t] \Psi^{*}[x_{10},x_{20},t]$. 

The probability of measuring the particle first and the mirror later is given by a product of two probabilities for one-body measurements. The first, the probability of measuring the particle around $ x_{10}$ at time $t_{10}$ without knowledge of the mirror's position, is determined from the average of the two-body PDF over the mirror coordinate $x_{2}$, $\textmd{Pr}_{I}[x_{10}-\Delta x_{1}<x_{1}<x_{10}+\Delta x_{1}]=\textmd{PDF}_{1}\Delta x_{1}$ with $\textmd{PDF}_{1}=\int \Psi[x_{10},t_{10},x_{2},t_{10}] \Psi^{*}[x_{10},t_{10},x_{2},t_{10}]dx_{2}$. 

The next, the probability of measuring the mirror at $x_{2}$ and $t_{2}$ after having measured the particle at $x_{10}$ within $\Delta x_{1}$ at $t_{10}$, is given by $\textmd{Pr}_{II}[x_{20}-\Delta x_{2}<x_{2}<x_{20}+\Delta x_{2}]=\textmd{PDF}_{2}\Delta x_{2}$ with $\textmd{PDF}_{2}= \Psi[x_{10},t_{10},x_{2},t_{2}] \Psi^{*}[x_{10},t_{10},x_{2},t_{2}]\Delta x_{1}$. The probability of first measuring the particle at $x_{10}$ and time $t_{10}$ and then the mirror at $x_{2}$ and time $t_{2}$ is therefore $\textmd{Pr}_{I}\textmd{Pr}_{II}$. 

The predictions for synchronous measurement presented below are plots of the two-body PDF as a function of $x_{1}$ and $x_{2}$ for snapshots at $t_{1}=t_{2}=t$. The probability of measuring both the particle and mirror at the same time in a small region of these plots is $\Psi[x_{1},x_{2},t] \Psi^{*}[x_{1},x_{2},t] \Delta x_{1} \Delta x_{2}$.

In the asynchronous predictions presented, it is assumed that the particle has been measured and therefore the plots are of the one-body PDF for the mirror as a function of both $x_{2}$ and the position $x_{10}$ and time $t_{10}$ at which the particle was measured. These are snapshots at $t_{2}$. The probability of measuring the mirror in a small region at time $t_{2}$, having already measured the particle at $x_{10}$ and $t_{10}$, is then $\Psi[x_{10},t_{10},x_{2},t_{2}] \Psi^{*}[x_{10},t_{10},x_{2},t_{2}] \Delta x_{2}$. From these figures, only the probability of measuring the mirror after the particle has been measured (not the probability of first measuring the particle {\em and} then the mirror) can be determined. Conservation of probability for this one-body mirror wavefunction is addressed in the appendix.

While the validity of this asynchronous model is limited by the assumption of no interaction after the first measurement it also makes subtle assumptions about collapse of a two-body wavefunction: a measurement of the particle substate influences neither the mirror substate at $t_{10}$ nor the subsequent evolution of the mirror substate apart from fixing the parameters of the particle substate in the two-body wavefunction. A more detailed discussion is found in subsection \ref{sec:measurementtheory}.

\subsection{Details of the asynchronous method}
\label{sec:non-simultaneous}

Before reflection an uncorrelated solution to the Schr\"odinger equation is given by
\begin{eqnarray}
\Psi_{0} \propto \exp[i (k x_{1}-\frac{\hbar k^{2}}{2m}t_{1}+K x_{2}-\frac{\hbar K^{2}}{2M}t_{2})].
\label{eq:ScheqnUnentangled}
\end{eqnarray}
An incident wavegroup constructed from this then leads to uncorrelated predictions about the probability of finding the particle at $(x_{1},t_{1})$ and mirror at $(x_{2},t_{2})$.

For the reflected wavefunction, the solution to eqn. \ref{eq:ScheqODE2} must vanish at $x_{1}=x_{2}$ to satisfy the boundary condition at the mirror and not exist for $x_{rel}<0$ (or $x_{1}>x_{2}$) since the particle cannot move through the mirror (for the uncorrelated incident state, however, there is no interaction and the particle does move past the mirror). 

In this transformed system, a solution to eqn. \ref{eq:ScheqODE2} can be constructed from the superposition of incident and ``reflected'' wavefunctions in the cm-rel system (in much the same way as is the solution for a wave traveling along a string toward a rigidly clamped boundary is constructed from free string solutions traveling in opposite directions),
\begin{equation}
\psi_{rel}=(e ^{i \phi_{in}}-e ^{i \phi_{ref}}) \theta [x_{rel}],
\label{eq:PsiSeparable}
\end{equation}
where $\theta [x_{rel}]$ is the unit step function. The only difference between the arguments of the two exponentials is the sign of the relative wavevector $K_{rel}$ corresponding to reflection in the relative coordinate. That is,
\begin{eqnarray}
\phi_{in/ref}= {\bm \pm}~ K_{rel} x_{rel}-\frac{\hbar K_{rel}^{2}}{2 \mu}t_{rel},
\label{eq:ScheqPhase}
\end{eqnarray}
where the initial velocities must allow reflection to occur. Relative and center of mass times $t_{rel}$ and $t_{cm}$ are introduced and associated with the relative and center of mass energies. These time variables satisfy the same properties as do $t_{1}$ and $t_{2}$ but in this case provide the notation needed in separating the energies associated with the relative and center of mass subsystems.

The solution to eqn. \ref{eq:ScheqODE1} for reflection is given by 
\begin{equation}
\psi_{cm}=e ^{i (K_{cm} x_{cm}-E_{cm} t_{cm}/\hbar }).
\label{eq:Psicm}
\end{equation}
The complete solution for an eigenstate of energy in reflection is then $\Psi[x_{cm},t_{cm},x_{rel},t_{rel}]=\psi_{cm}\psi_{rel}$.

The particle-mirror system has now been partitioned into separable center of mass and relative coordinate subsystems.  This separable solution, for the two uncorrelated substates, can be used to construct a wavegroup whose substates yield uncorrelated wavegroups associated with the cm and relative coordinates which satisfies initial and/or boundary conditions. A measurement of the cm position affects neither the time evolution of the wavegroup associated with the relative motion nor introduces any correlation between the cm and relative positions.

While the energy and momentum of the center of mass subsystem are unaffected by reflection, that is not the case for the relative subsystem where the relative wavevector changes sign upon reflection. Therefore, in this separable system interference of incident and reflected wavefunctions, determined by the PDF
\begin{eqnarray}
\Psi \Psi^{*}=4 \sin^{2}[K_{rel}x_{rel}],
\label{eq:SeparableInterference}
\end{eqnarray}
is associated only with the relative coordinate subsystem.

The change from cm-rel to particle-mirror systems is accomplished by using the following relations in the separable solutions given by equations \ref{eq:Schtot}, \ref{eq:PsiSeparable}, \ref{eq:ScheqPhase}, and \ref{eq:Psicm}: $K_{cm}=k+K$, $K_{rel}=(Mk-mK)/M_{tot}$, $x_{rel}=x_{1}-x_{2}$, $x_{cm}=(mx_{1}+Mx_{2})/M_{tot}$, $E_{rel}=\hbar^2K_{rel}^{2}/2\mu$, and $E_{cm}=\hbar^2K_{cm}^{2}/2(m+M)$. 

These relations, however, do not address the two-time labeling issue in the particle-mirror system. To do so, note that the energy of the reflected particle, given by $p_{ref}^{2}/2m$ with $p_{ref}=\hbar \partial \phi_{ref}/\partial x_{1}$, is associated with the temporal coordinate $t_{1}$. Similarly, the energy for the mirror is $P_{ref}^{2}/2M$ with $P_{ref}=\hbar \partial \phi_{ref}/\partial x_{2}$ and is associated with the temporal coordinate $t_{2}$. Both of these energies and momenta are consistent with those of a classical particle reflecting from a moving mirror. These are manifest in the two-body wavefunctions, however, as a Doppler shift.

Application of these transformation relations then changes $e ^{i \phi_{in}}$ in equation \ref{eq:PsiSeparable} into equation \ref{eq:ScheqnUnentangled}. The two-time expression for  $e ^{i \phi_{ref}}$ in the particle-mirror system, although simple to calculate using the procedure just outlined, is too large to present here. This complexity is a consequence of the correlations generated in reflection.

Correlated interference of these incident and reflected wavefunctions in the particle and mirror subsystems is then given by (transforming eqn. \ref{eq:SeparableInterference})
 \begin{eqnarray}
\Psi \Psi^{*}=4 \sin^{2}[(mK-Mk)\{(m+M)(x_{1}-x_{2})\notag \\
-\hbar (k+K)(t_{1}-t_{2}\}/(m+M)^2].
\label{eq:EntangledInterference}
\end{eqnarray}
This is similar to Gottfried's joint or simultaneous PDF for the interference obtained in the correlation between two particles produced in a momentum-conserving decay after each has traversed separate double slits \cite{Gottfried} when $t_{1}=t_{2}$. Note also that interference of the mirror and particle are coupled in eqn. \ref{eq:EntangledInterference}, illustrating how many-body systems interfere with `themselves' rather than only the particle and mirror each interfering with itself (in which case there would be no correlation) \cite{silverman}.

To gain familiarity with this result, consider a simultaneous measurement. For fixed  $x_{1}$, the approximation $m/M<<1$ leads to interference for the mirror which varies from maximum to minimum through a distance
\begin{eqnarray}
\Delta x_{2} \approx  \pi \hbar/(m(v-V)).
\label{eq:fringespacing}
\end{eqnarray}
Similarly for fixed  $x_{2}$ this approximation leads to interference for the particle which varies from maximum to minimum through a distance $\Delta x_{1}=\Delta x_{2}$. For $V=0$ both the mirror and particle fringes are spaced at half the deBroglie wavelength of the {\em particle}, which can be up to $10^{-6}$ m for ultra cold atoms \cite{cronin}.

The time dependence in equation \ref{eq:EntangledInterference} is determined by the time components of the phase of the incident wavefunction, $\Phi_{in}[t_{1},t_{2}]=p_{in}^{2}t_{1}/2m + P_{in}^{2} t_{2}/2M$,  and the reflected wavefunction, $\Phi_{ref}[t_{1},t_{2}]=p_{ref}^{2}t_{1}/2m + P_{ref}^{2} t_{2}/2M$. The temporal part of the joint PDF depends on $\Phi_{in}[t_{1},t_{2}]-\Phi_{ref}[t_{1},t_{2}]$. For simultaneous measurements ($t_{1}=t_{2}=t$) this phase difference is zero since the time variable factors from all energy terms and the total energy before and after reflection does not change: $\Phi_{in}[t]-\Phi_{ref}[t]=(p_{in}^{2}/2m + P_{in}^{2}/2M-p_{ref}^{2}/2m - P_{ref}^{2}/2M) t=0$. That, however, is not the case for non-simultaneous measurements since the times of measurement of the particle and mirror differ and are therefore no longer a common factor of all energy terms in the phase. This leads to the Doppler effect which is described next.

Consider an ensemble of identically prepared particle-mirror systems. Let a measurement on every member of the ensemble be made at both fixed particle-mirror positions, $x_{1}$, $x_{2}$, and time $t_{2}$ while the particle is measured at different times $t_{1}$ for different members of the ensemble. The time dependent interference pattern from eqn. \ref{eq:EntangledInterference} emerges from these ensemble measurements as the expected ``beat frequency" $\Omega$ \cite{beat} associated with interference of the incident and reflected particle substates (the superposition of states with different energies commensurate with the energy exchanged in reflection). If instead, $x_{1}$, $x_{2}$, and $t_{1}$ are fixed while the mirror position is measured at different times $t_{2}$, the ``beat frequency" is that associated with superposing mirror substates differing in energy. These particle and mirror beat frequencies are identical due to the same energy being exchanged between the particle and mirror in reflection and are given by
\begin{eqnarray}
\Omega=\frac{mM(v-V)(mv+MV)}{\hbar(m+M)^{2}}.
\label{eq:beat}
\end{eqnarray}

\subsection{Wavegroups: Simultaneous measurement}
\label{sec:wavegroupsimultaneous}

To better understand the experimental consequences of these results, wavegroups are next formed from a superposition of the incident and reflected `energy eigenstates' (given by eqn. \ref{eq:PsiSeparable}) expressed in terms of the correlated particle and mirror substates rather than the cm and relative substates. It is assumed that the initial particle and mirror Gaussian substates are sufficiently spatially separated that any probability of the particle initially being on the ``wrong'' side of the mirror is negligible. An analytic expression for such wavegroups can be obtained for a Gaussian distribution in wavevector components $k$ and $K$ (or velocities $v$ and $V$). For the mirror this is proportional to $\exp [-(K-K_{0})^{2}]/(2 \Delta K^{2})$ where the peak of the distribution is at $K_{0}$ and $\Delta K$ is its width while for the particle this is proportional to $\exp [-(k-k_{0})^{2}]/(2 \Delta k^{2})$ where the peak of the distribution is at $k_{0}$ and $\Delta k$ is its width.

\begin{center}
\begin{figure}
\includegraphics[scale=0.29]{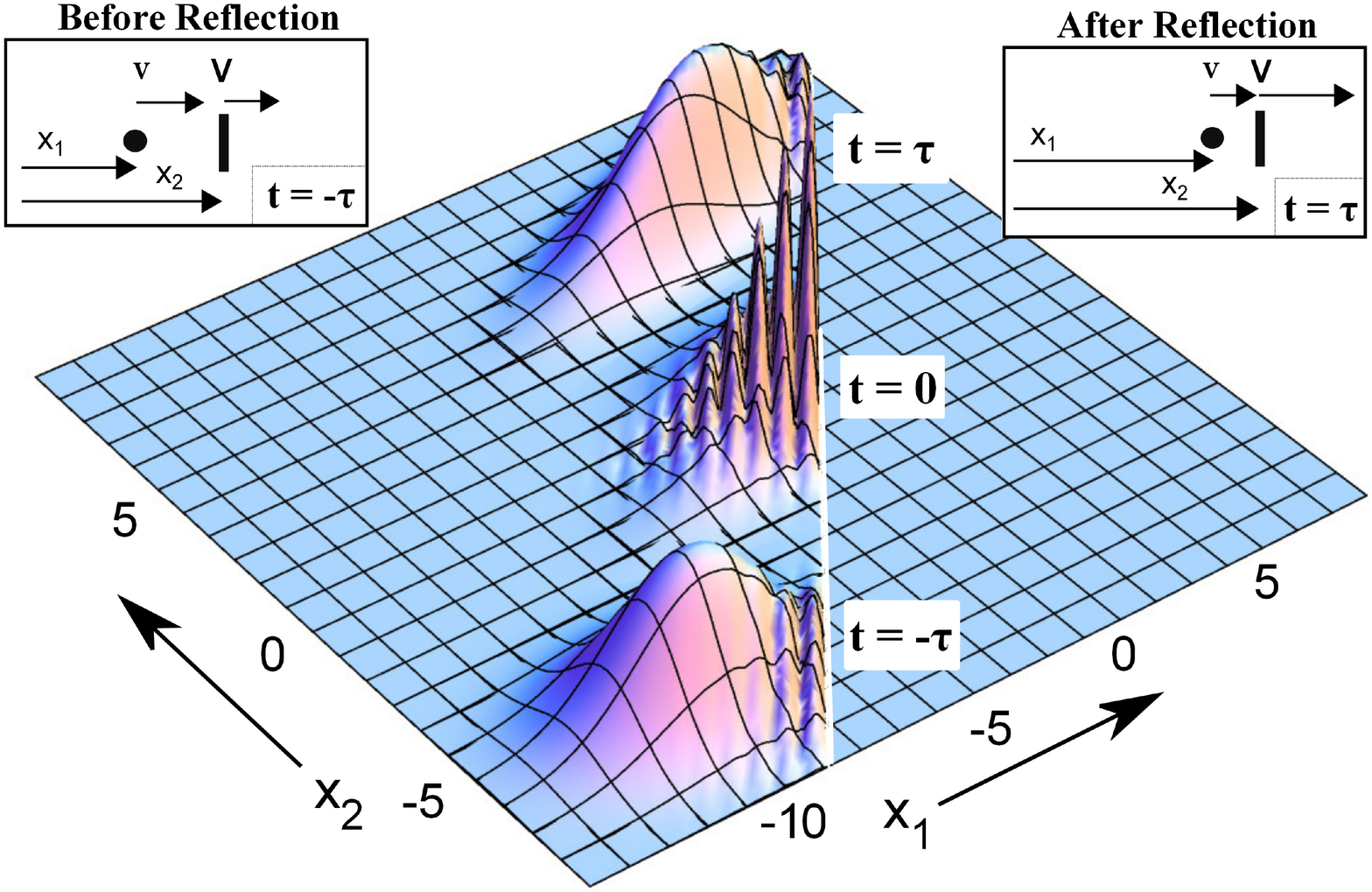}
\caption{Two-body simultaneous joint probability density snapshots for three sequential times vs coordinates $(x_{2},x_{1})$ for a particle reflecting from a mirror. The lower PDF waveform moves toward the diagonal white line, corresponding to $x_{1}=x_{2}$, then reflects in the middle snapshot where the incident and reflected two-body wavefunctions `overlap', and finally it moves away from the diagonal in the upper snapshot. The correlated interference fringes are spaced by about half the deBroglie wavelength of the {\em particle}. The upper left inset is a schematic of the `classical' analog before reflection while the upper right inset is that after reflection with initial and final particle and mirror velocities $v$, $V$, $v_{f}$, and $V_{f}$ respectively. There is no classical analog for the middle snapshot.}
\label{fig:interference}
\end{figure}
\end{center}

Consider first {\em simultaneous} measurement of the particle and mirror. In fig. \ref{fig:interference} snapshots of such a two-body joint PDF are shown at three times, $t_{1}=t_{2}=(-\tau, 0, \tau)$, for $M/m=100$, $\Delta K/\Delta k=2$, and $K/k=60$. The incident wavegroup propagates in the $(x_{1},x_{2})$ plane along a line whose slope is determined by a ratio of the group velocities of each substate and spreads due to dispersion independently in each direction. The particle and mirror initially move in the positive direction for the parameters chosen. During reflection the incident and reflected wavegroups overlap resulting in interference. After reflection the speed of the mirror increases while the particle continues moving in the positive direction with decreased speed (see the insets for the classical analog). Careful inspection of the lower and middle snapshots just to the left of the diagonal white line  confirms the prediction that the spatial location of the interference maxima and minima do not depend on time as given in eqn. \ref{eq:EntangledInterference} when $t_{1}=t_{2}$. Wavegroup distortion shown in the upper reflected snapshot is discussed below. In the cm-rel system (not shown) the ``fringes" are aligned parallel to the cm coordinate illustrating the result given in eqn. \ref{eq:SeparableInterference}.

\begin{center}
\begin{figure}
\includegraphics[scale=0.3]{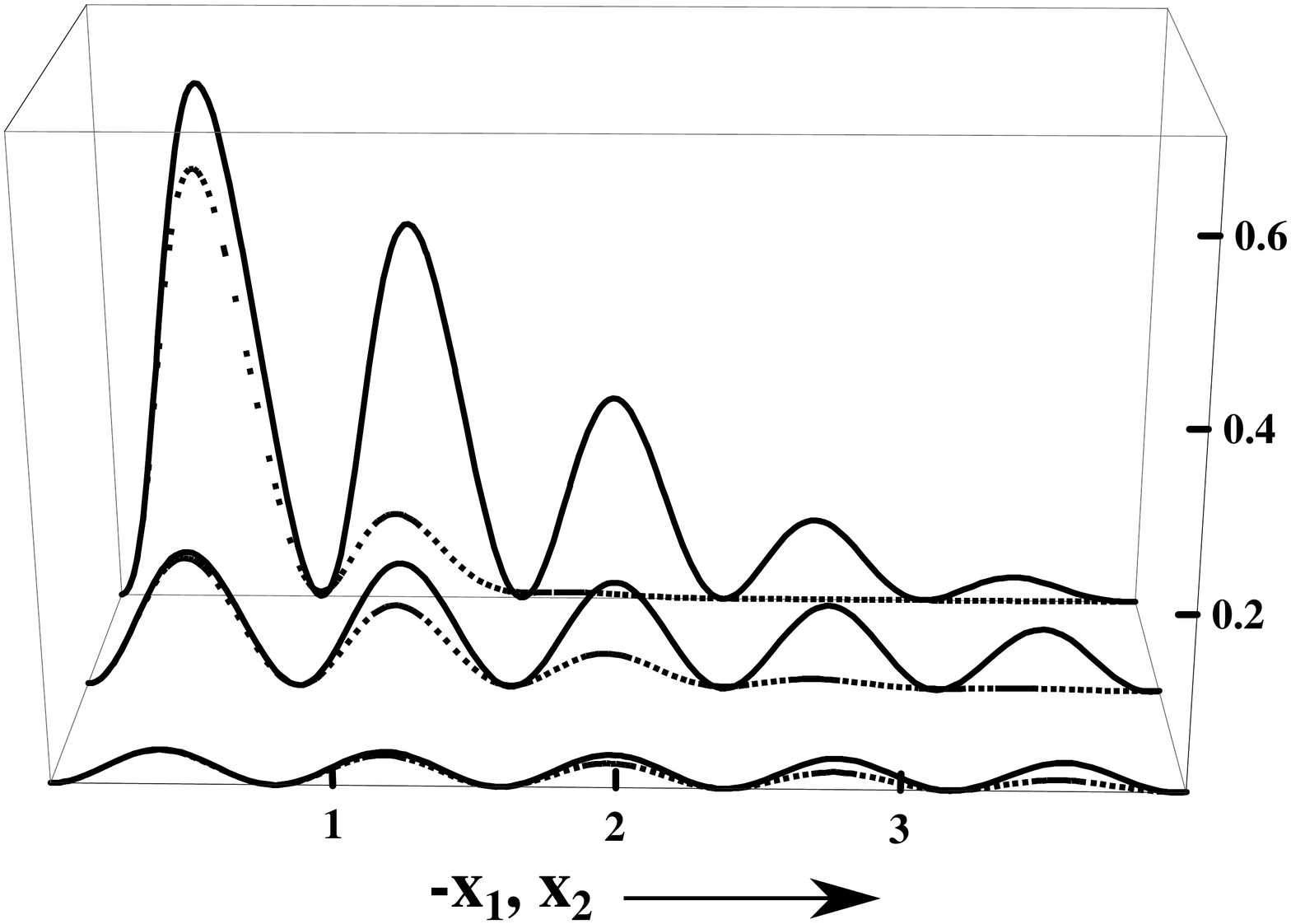}
\caption{Slices of a snapshot similar to the middle one from fig. \ref{fig:interference} which show the fringe spacing along the $x_{2}$ axis for $x_{1}=0$ (dashed lines) and along the $x_{1}$ axis for $x_{2}=0$ (solid lines). The $x_{1}$ axis has been inverted to display both the dashed and solid lines together. Although each graph has $\Delta K/\Delta k=2$ the value of $\Delta K$ increases sequentially by a factor of $2$ from the front to the back of the figure. }
\label{fig:fringespacing}
\end{figure}
\end{center}

A slice of fig. \ref{fig:interference} for $x_{1}=0$ along the $x_{2}$ coordinate is shown in fig. \ref{fig:fringespacing} (the solid line) along with a slice of this figure for $x_{2}=0$ along $x_{1}$ (the dashed line) for different bandwidth wavegroups. This demonstrates essentially the same fringe spacing for the particle and mirror substates with narrow bandwidth wavegroups, as discussed for the approximation $M/m>>1$ following eqn. \ref{eq:fringespacing}.

The minima of the interference shown in fig. \ref{fig:interference} correspond to positions where the particle and mirror can never simultaneously be found. Verification of this result requires simultaneous cm measurement of the particle and mirror with instruments which have a spatial resolution that is smaller than this fringe spacing along both coordinates.

\subsection{Wavegroups: fixed $\bf{t_{10}}$ while $\bf{t_{2}}$ varies}
\label{sec:non-simultaneous limit}

Next consider non-simultaneous measurements of the particle at position $x_{10}$ and time $t_{10}$ while allowing $t_{2}$ to vary until a measurement is made of the mirror's cm position. This can be categorized into two regimes. Measurement of the particle either occurs in the region where the incident and reflected wavegroups do not overlap, regime A, or in the region where the interference similar to that shown in fig. \ref{fig:interference} occurs, regime B.

To illustrate such asynchronous measurements, the particle is measured at a particular $x_{10}$ and $t_{10}$ while the mirror substate then evolves with $\Psi[x_{10},t_{10},x_{2},t_{2}]$. This is shown as one-body PDF plots of snapshots at different times $t_{2}$ using a 3-D graph of $x_{2}$ vs fixed values of $x_{10}$ and $t_{10}$.

\subsubsection{Regime A: the particle is measured when there is no incident and reflected wavefunction overlap}
\label{regime A}

\begin{center}
\begin{figure}
\includegraphics[scale=0.25]{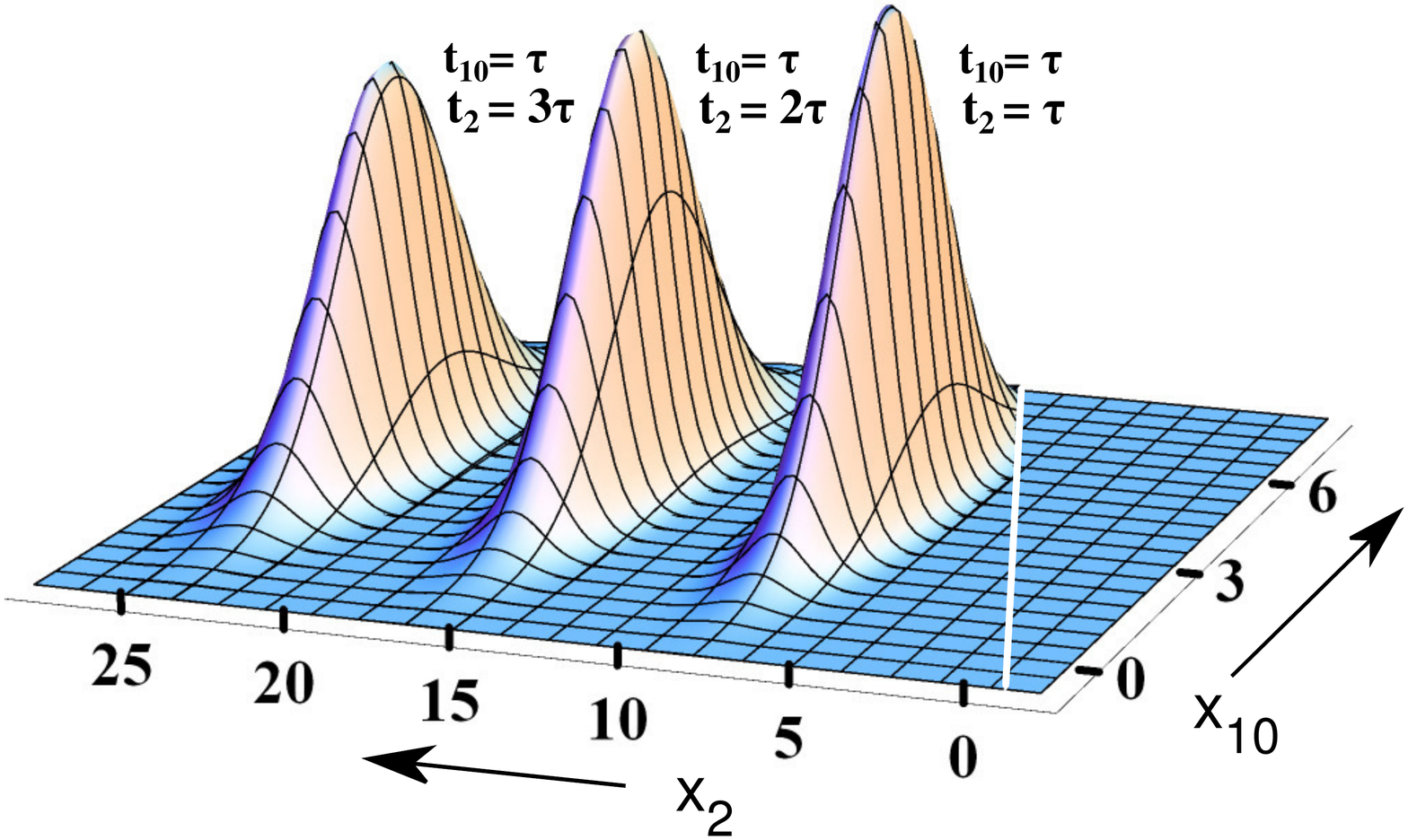}
\caption{Regime A one-body PDF plots for the mirror at $t_{2}=\tau, 2\tau, 3\tau$ when the particle has been measured at $t_{10}=\tau$ and $x_{10}$. Since $t_{2}=t_{10}=\tau$ for the rightmost PDF, it is similar to the upper snapshot in fig. \ref{fig:interference}. The diagonal white line is the same as that in fig. \ref{fig:interference}.}
\label{fig:regimeA}
\end{figure}
\end{center}

All three joint PDF snapshot results for regime A shown in fig. \ref{fig:regimeA} are for $t_{10}=\tau$. The one farthest to the right is similar to the upper PDF waveform snapshot in fig. \ref{fig:interference} since it occurs at $t_{2}=t_{10}=\tau$. The other two snapshots are for times, $t_{2}= 2\tau$ and $3\tau$ with $M/m=3$, $\Delta K/\Delta k=2$, and $K/k=1.8$. Note that this mirror wavegroup moves only along the $x_{2}$ axes and disperses.

It is useful to compare the physical interpretation of the joint PDF shown in fig. \ref{fig:interference} with that of the one-body PDF of fig. \ref{fig:regimeA}. In the former case the probability of finding the particle and cm of the mirror in a region centered around $x_{1}=a$ and $x_{2}=b$ simultaneously at time $t$ is given by $\int^{b+\delta b}_{b-\delta b}\int^{a+\delta a}_{a-\delta a} PDF[x_{1},t,x_{2},t] dx_{1} dx_{2}$. In fig. \ref{fig:regimeA} the particle is measured at $t_{10}=\tau$ while the snapshots are given for different values of when and where the cm of the mirror is measured. For the leftmost PDF in this figure the probability of measuring just the mirror, once the particle has been measured, is given by $\int^{b+\delta b}_{b-\delta b} PDF[x_{10},\tau,x_{2},3\tau] dx_{2}$.

\subsubsection{Regime B: the particle measured in the overlap region}
\label{regime B}

\begin{center}
\begin{figure}
\includegraphics[scale=0.28]{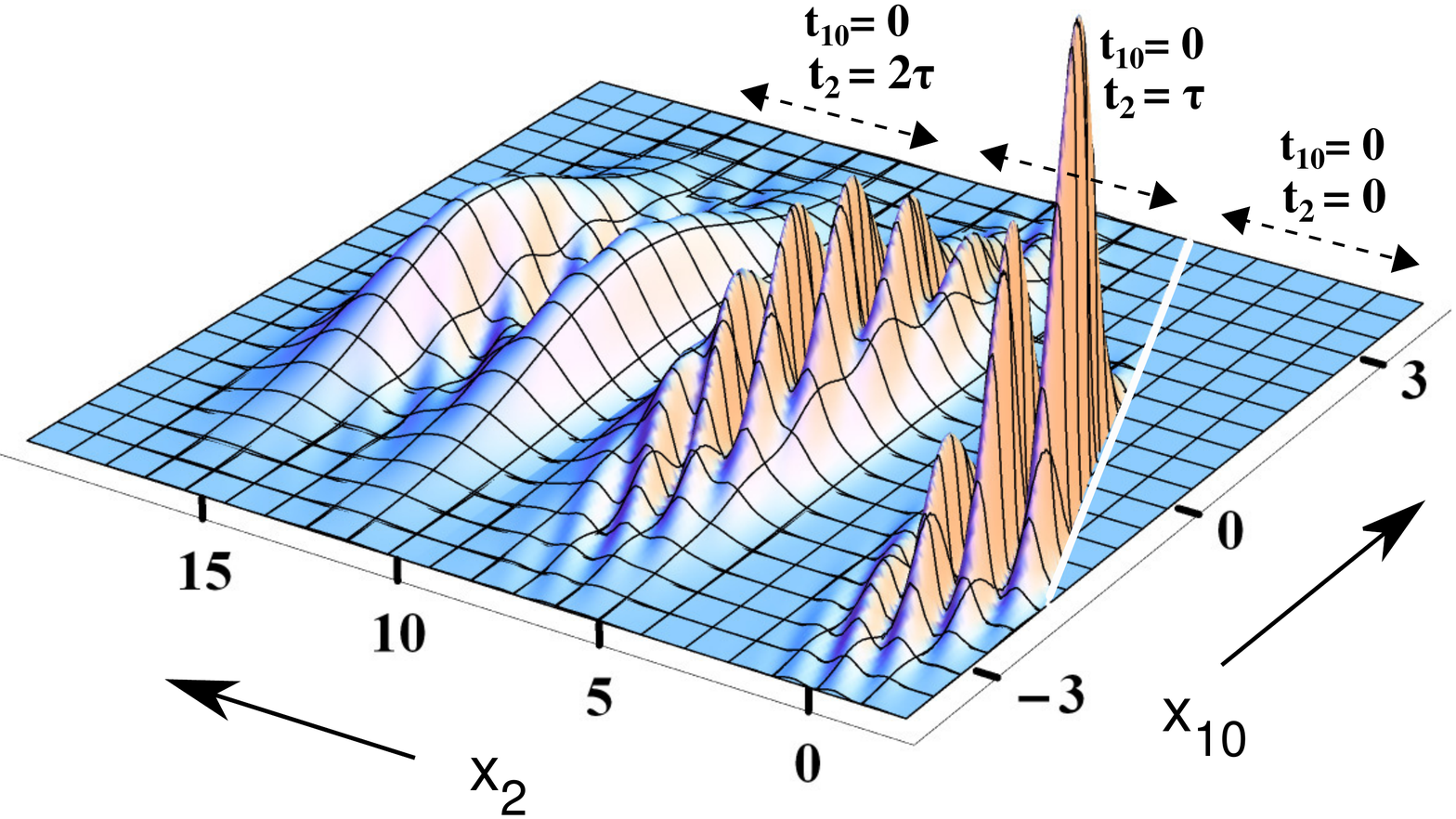}
\caption{Regime B one-body mirror PDF plots for $t_{10}=0$ and 3 sequential times, $t_{2}=0, \tau, 2\tau$. The rightmost PDF waveform is similar to the middle PDF waveform in fig. \ref{fig:interference}. The diagonal white line is the same as that in fig. \ref{fig:interference}. This is a PDF illustration of the splitting of the mirror states shown schematically in fig. \ref{fig:overview}}
\label{fig:regimeB}
\end{figure}
\end{center}

All three joint PDF snapshot results for regime B, shown in fig. \ref{fig:regimeB}, are for $t_{10}=0$. The one farthest to the right is similar to the middle PDF waveform snapshot in fig. \ref{fig:interference} since it occurs at $t_{2}=t_{10}=0$. The middle snapshot of fig. \ref{fig:regimeB} occurs predominately between $x_{2}=4$ and $9$ for $t_{2}= \tau$, while the leftmost snapshot consists of two ``bumps" encompassing the region between $x_{2}=9$ and $18$ for $t_{2}= 2 \tau$. As in the previous figure $M/m=3$, $\Delta K/\Delta k=2$, and $K/k=1.8$.

The complexity of the joint PDF in fig. \ref{fig:regimeB}, in comparison with that of regime A, is a consequence of measuring the particle in the interference region. That is, there are two indistinguishable ways that the particle could have reached the interference region. It could have come from the incident {\em or} reflected particle wavegroup substates. This lack of knowledge about the particle is manifest in the subsequent mirror wavegroup which then consists of a superposition in which the mirror has yet to reflect {\em and} has already reflected the particle. 

These two mirror states have different speeds due to the mirror recoil in one but not the other. To see this one need only change the speed of the incident particle wavegroup. The speed of the mirror state which has not reflected the particle does not change while the speed of the mirror state which reflected the particle increases. Overlap of these two mirror wavegroup states is shown in fig. \ref{fig:regimeB} from complete overlap (right hand snapshot) to partial overlap (middle snapshot) and finally to virtually complete separation (left hand snapshot) due to the differing speeds of the two mirror states.

Although these results illustrate important issues in non-simultaneous measurement, they have explored only a limited set of parameters and interferometric geometries. Rather than present a comprehensive treatment, we use three examples to further probe the consequences of asynchronous measurements. First, the beamsplitting effect, that a measurement on the particle has on the mirror, is explored in more depth using a particle whose mass is equal to that of the mirror. Second, a manifestation of the Doppler effect is illustrated, using a time scale shorter than that used in the figures above. Finally, a specific example of reflection of a microscopic particle from a macroscopic mirror is described.

\subsubsection{Regime B: measurement functioning as a beamsplitter using a coherence transfer example}
\label{subsec:beamsplitter}

\begin{center}
\begin{figure}
\includegraphics[scale=0.30]{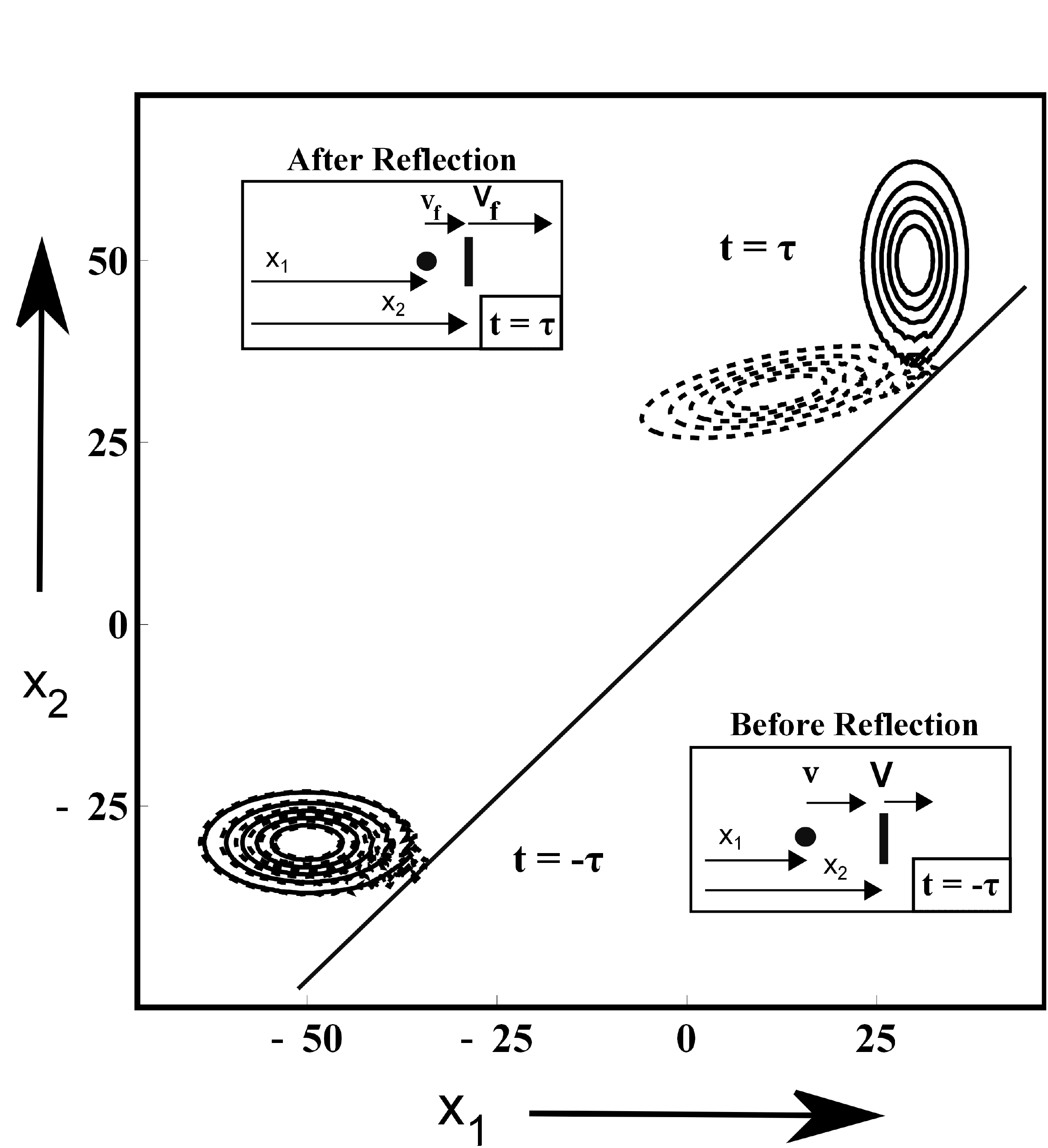}
\caption{Contour plots of the joint probability density snapshot for simultaneous measurement of a particle reflecting from a mirror, similar to fig. \ref{fig:interference} but without the interference or ``overlap'' region snapshot, illustrating coherence transfer for two different particle-mirror mass ratios. The solid and dashed lines are for a spread in velocities given by $\Delta V/\Delta v=10$ with $M/m=1$ and $M/m=20$, respectively. Note the exchange of wavegroup widths between particle and mirror substates for $M/m=1$. The diagonal white line is the same as that in fig. \ref{fig:interference}.}
\label{fig:coherencetransfer}
\end{figure}
\end{center}

After reflection, the spatial width of the mirror wavegroup substate is exchanged with that of the particle wavegroup substate when $M=m$. This is most easily seen by constructing a particle-mirror wavefunction with different bandwidths for the particle and mirror wavegroup substates and is shown in fig. \ref{fig:coherencetransfer}, which is a contour plot of joint PDFs similar to figure \ref{fig:interference}, but without the snapshot in the interference region. The solid and dashed contours correspond to $M/m=1$ and $M/m=20$ respectively with the spread in velocities given by $\Delta V/\Delta v=10$.

This result can be understood by comparing classical and quantum reflection. In a one-dimensional classical collision, conservation of energy and momentum result in the exchange of particle-mirror velocities independent of either velocity for $m=M$. This is manifest quantum mechanically in the exchange of commensurate  substate parameters $k$ and $K$ between the incident and reflected two-body wavefunctions. If an incident particle substate, consisting of only one harmonic component (corresponding to speed $v$) reflects from a mirror substate with many velocity components, then each harmonic component of the mirror substate (corresponding to different values of $V$) reflects the particle substate  and therefore acquires velocity $v$, while the reflected particle substate acquires different velocity values for each reflected component of the mirror wavegroup. This results in the reduction of the mirror bandwidth and an increase in the particle bandwidth, which is manifest in fig. \ref{fig:coherencetransfer} as the exchange of incident and reflected wavegroup shapes. It also is responsible for the distortion of the reflected wavegroup shape in fig. \ref{fig:interference}.

\begin{center}
\begin{figure}
\includegraphics[scale=0.3]{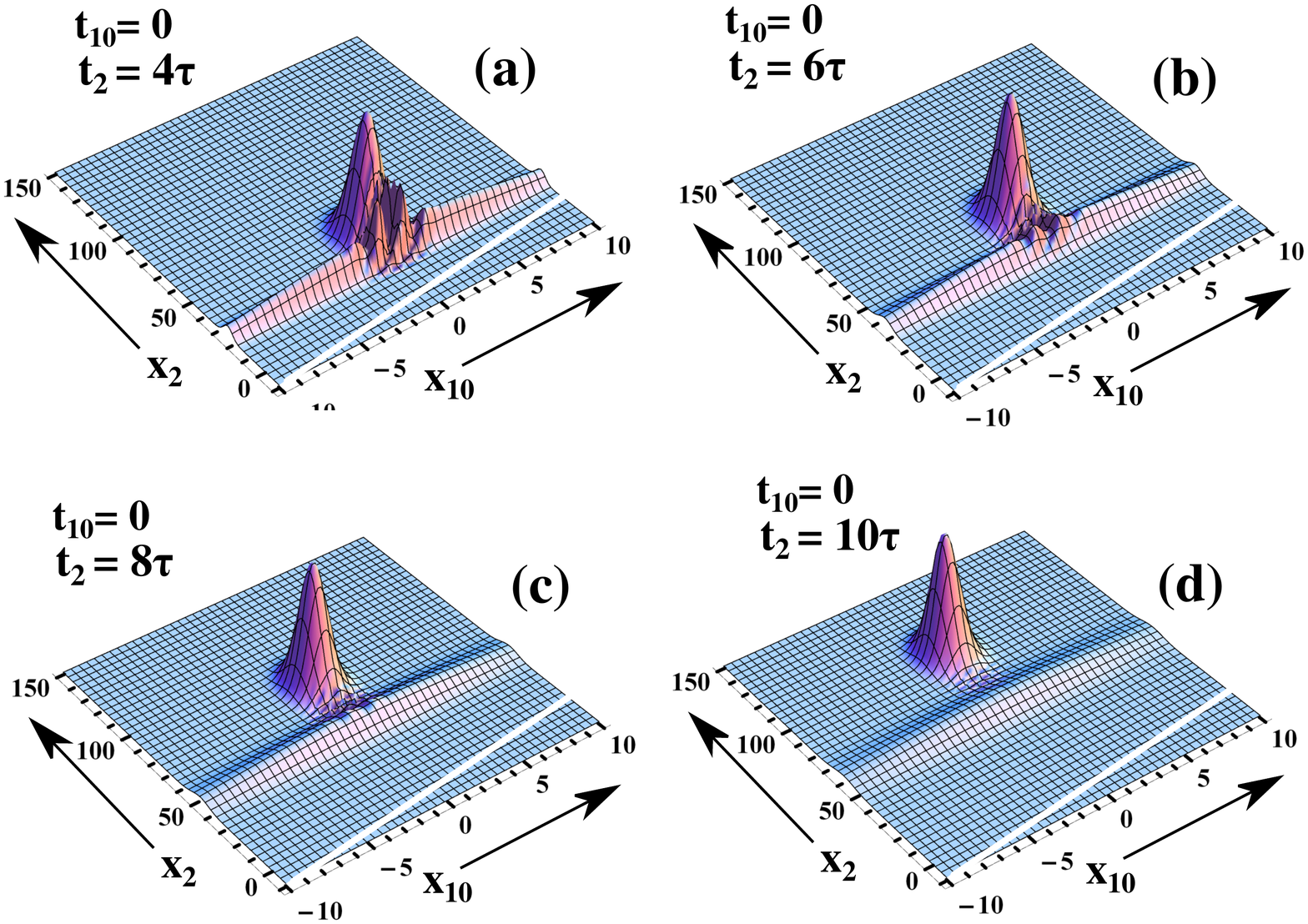}
\caption{One-body mirror PDFs for the same equal mass parameters as in fig. \ref{fig:coherencetransfer} except that the particle was measured at $x_{10}$ and $t_{10}=0$ while the mirror is measured at $x_{2}$ with incremental increases in the time $t_{2}$. Note the difference in scale along the axes. The diagonal white line is the same as that in fig. \ref{fig:coherencetransfer}. The distinct mirror substates in part (d) correspond to those which either have or have not reflected the particle. This is another PDF illustration of the splitting of the mirror states shown schematically in fig. \ref{fig:overview}}
\label{fig:2TequalMass}
\end{figure}
\end{center}

Consider next non-simultaneous joint PDFs for equal particle and mirror masses. However, rather than choosing measurement of the particle at $t_{10}=\pm \tau$, as shown in fig. \ref{fig:coherencetransfer}, the particle is measured at $t_{10}=0$ to illustrate asynchronous measurement in the interference region. The results, shown in fig. \ref{fig:2TequalMass}, again illustrate splitting of the mirror. However, these states differ dramatically in their shapes and in their splitting ratios. That is, depending on the position that the particle is measured at, a slice of the joint PDF along the $x_{2}$ axis can reveal one or two peaks of unequal height. The long PDF waveform parallel to the $x_{10}$ axis is that of the mirror substate associated with no reflection, while the narrow peak corresponds to the mirror substate which reflected the particle. The differing speeds of these two waveforms, due to mirror recoil in one but not the other, are evident in the figure.

The time order of measuring the particle and then the mirror can be reversed. In regime B, the splitting then generates two states consisting of a superposition in which the particle has yet to reflect {\em and} has already reflected from the mirror.

\subsubsection{Regime B: measurement of the particle resulting in a beat frequency for the mirror}
\label{subsec:beat}

\begin{center}
\begin{figure}
\includegraphics[scale=0.28]{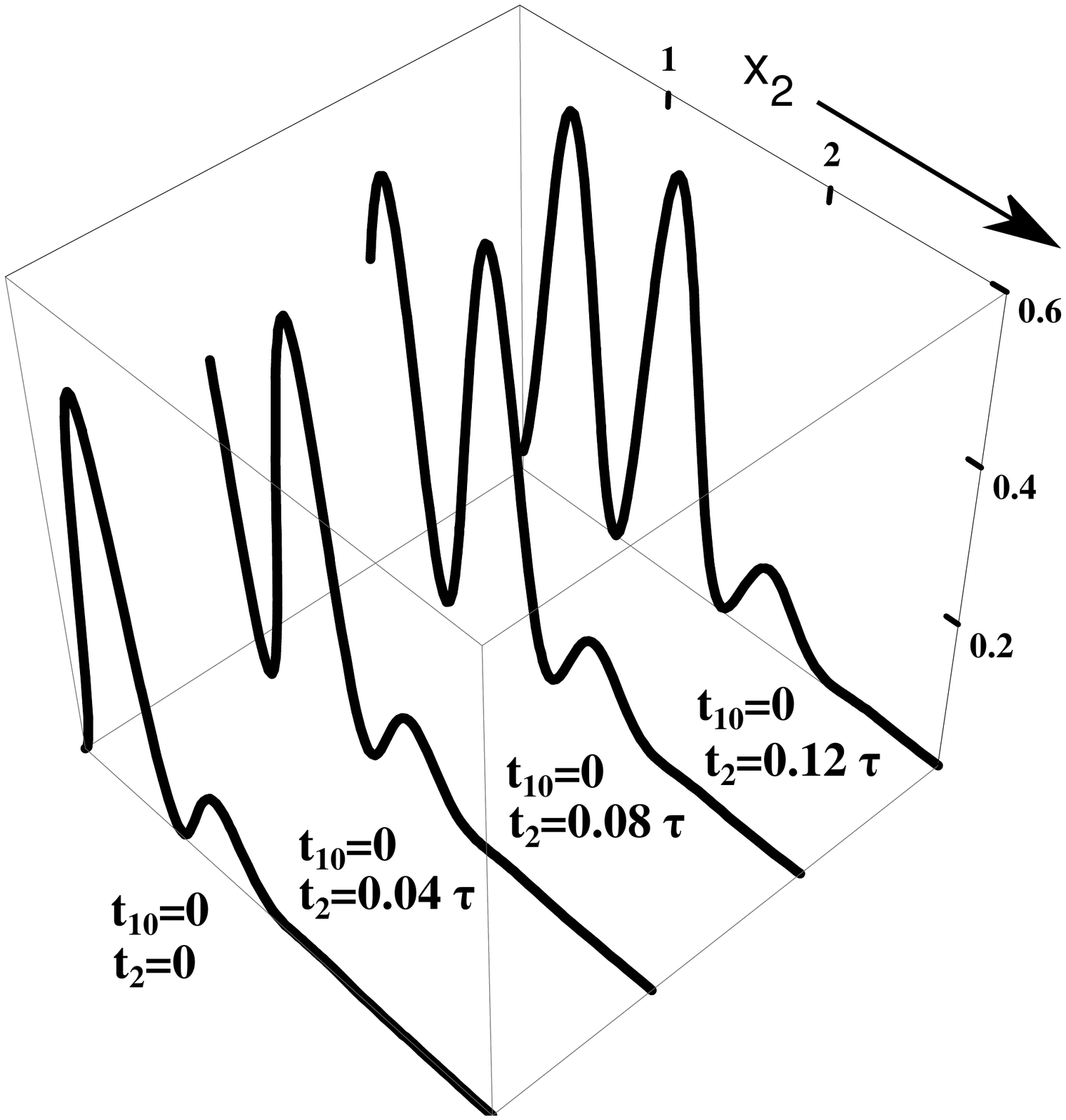}
\caption{One-body PDF snapshots of the mirror after the particle was measured at $x_{10}=0$ and $t_{10}=0$, starting at time $t_{2}=0$ (which is a slice of fig. \ref{fig:interference}) as the leftmost plot. The remaining plots to the right progressively increase the time $t_{2}$ when the mirror is measured by $0.04 \tau$.}
\label{fig:doppler}
\end{figure}
\end{center}

Variation of $t_{2}$, on a temporal scale smaller than that shown in fig. \ref{fig:regimeB}, reveals the energy differences between the superposition of these two mirror substates, in regime B, as a ``beat frequency'' in the joint PDF. This is related to the result for harmonic states given in eqn. \ref{eq:beat} and is illustrated for wavegroups in fig. \ref{fig:doppler}. A slice, taken from fig. \ref{fig:interference}, along the $x_{2}$ axis for $x_{10}=0$ and  $t_{10}=t_{2}=0$, is shown in fig. \ref{fig:doppler} as the leftmost plot. Plots to the right of this are shown at $x_{10}=0$ and $t_{10}=0$ for $t_{2}= 0.04 \tau,~0.08 \tau$, and $0.12 \tau$, respectively. All other parameters are the same as used in fig. \ref{fig:interference}. This illustrates the expected beat frequency from the ``Doppler shift'' in reflection, although it is shrouded in two-body interference. Equation \ref{eq:EntangledInterference} predicts no such beat frequency for simultaneous measurements ($t_{1}=t_{2}$), which is discussed in more detail in subsection \ref{sec:asynchDoppler}.

\subsubsection{Regime B: interference for a mirror of mesoscopic/macroscopic mass}
\label{subsec:macroscopic}

One constraint for regime B is that the fringe visibility function must be non-zero and the incident and reflected particle-mirror wavegroups must `overlap' in the $(x_{1},x_{2})$ plane. The interference fringes are then determined predominately by a superposition of `energy eigenstates' \cite{Hamilton}. For example, the interference shown in fig. \ref{fig:interference} is determined predominately by eqn. \ref{eq:EntangledInterference}, when the wavegroups `overlap' in the center snapshot, where the longitudinal coherence lengths for both the particle and mirror are greater than the fringe spacing. In the upper snapshot of fig. \ref{fig:interference} there is neither `overlap' nor such interference.

The fringe ``visibility function'' is non-zero if each wavegroup substate `overlaps' within approximately a coherence length \cite{coherencelength}, which is given by $l_{c} \approx \lambda^{2}/\Delta \lambda = \lambda V/\Delta V$ \cite{hasselbach}. For particle substates, this can be $l_{c}^{particle}=10000$ \AA~for ultracold atoms \cite{cronin} or $l_{c}^{particle}=790$ \AA~for slow neutrons \cite{pushin}.

This longitudinal mirror coherence length can be estimated from the uncertainty in the mirror velocity. If it is in thermal equilibrium with the environment then $\Delta V_{thermal} \approx \sqrt{2k_{B}T/M}$, yielding $l_{c}^{thermal} \approx h/\sqrt{2Mk_{B}T}$, which is consistent with results for atoms in a Bose-Einstein condensate \cite{cronin}.

\begin{center}
\begin{figure}
\includegraphics[scale=0.28]{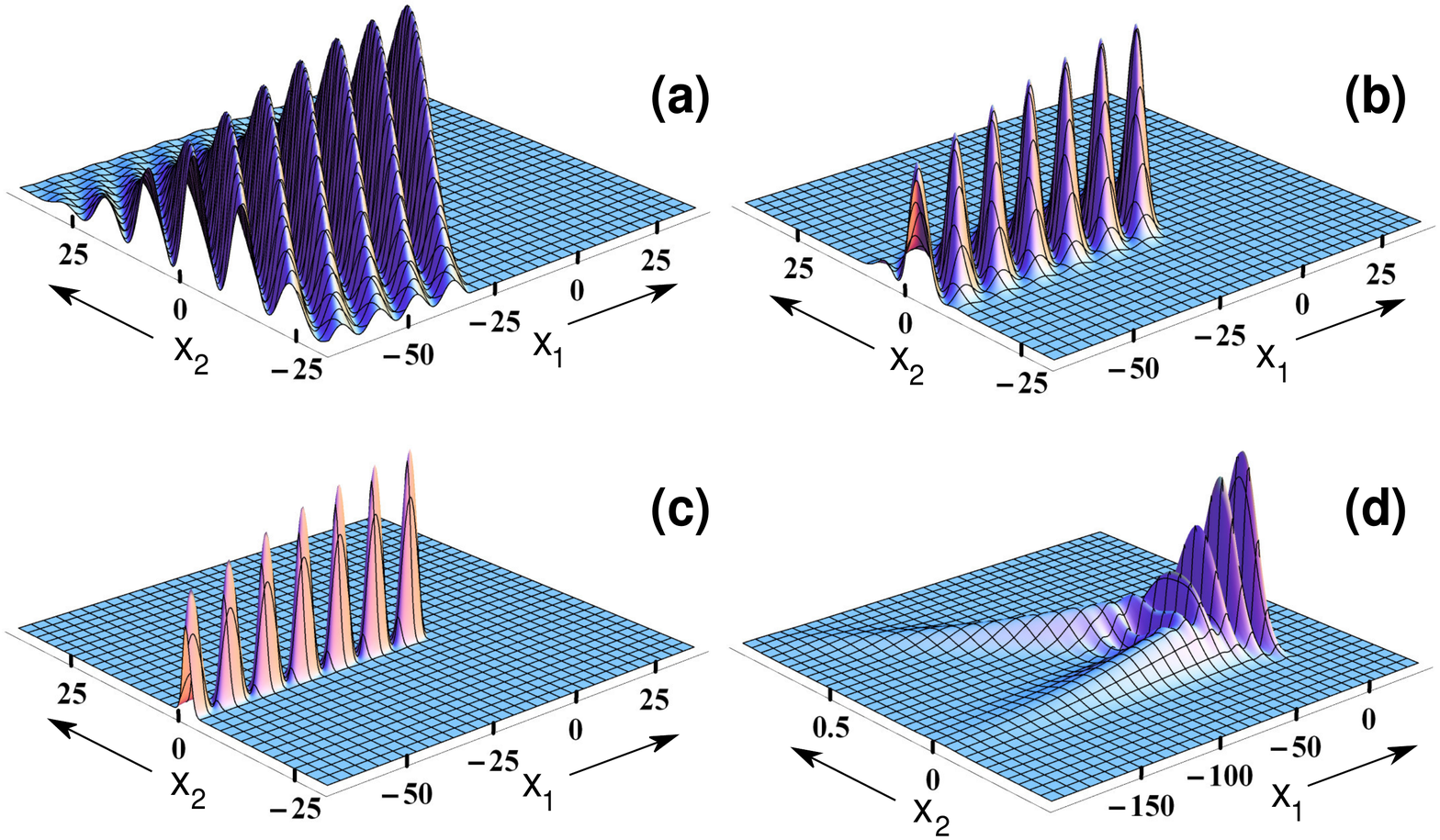}
\caption{Two-body PDF plots for simultaneous measurement similar to the center snapshot of fig. \ref{fig:interference}, but with different spreads in mirror velocities, for a fixed spread in particle velocity while all other parameters are the same. $\Delta V/\Delta v=80, 20, 5$, and $0.4$ for $M/m=200$ in parts (a), (b), (c), and (d) respectively. Note the change of scales in graph (d).}
\label{fig:spreadvelocity}
\end{figure}
\end{center}

Fig. \ref{fig:spreadvelocity} illustrates how variation in the longitudinal coherence length of the mirror substate affects the particle-mirror interference, for fixed particle coherence length and {\em simultaneous} measurements. Part  (a) shows a longer mirror coherence length than is used in fig. \ref{fig:interference}, while parts  (b) through (d) progressively reduce the coherence length of only the mirror substate. In fig. \ref{fig:spreadvelocity} (d), the coherence length of the mirror is so small that overlap is prevented, over a range of $x_{1}$ values where it was present before. Nevertheless, a slice along the $x_{2}$ axis for measurement of the particle at a particular $x_{1}$, indicates a splitting of the mirror substate into two states which do not overlap and are therefore distinguishable. This is a consequence of two ways that the particle could have reached $x_{1}$: it could have come from the incident {\em or} reflected particle wavegroup substates, due to the large particle coherence length. As the mirror's coherence length increases, these two ways overlap and generate correlated interference as shown in fig. \ref{fig:spreadvelocity} (a). As the mirror's coherence length decreases, the position of the mirror before reflection is distinguishable from that after reflection, resulting in no interference, as shown in part of fig. \ref{fig:spreadvelocity} (d).

\begin{center}
\begin{figure}
\includegraphics[scale=0.3]{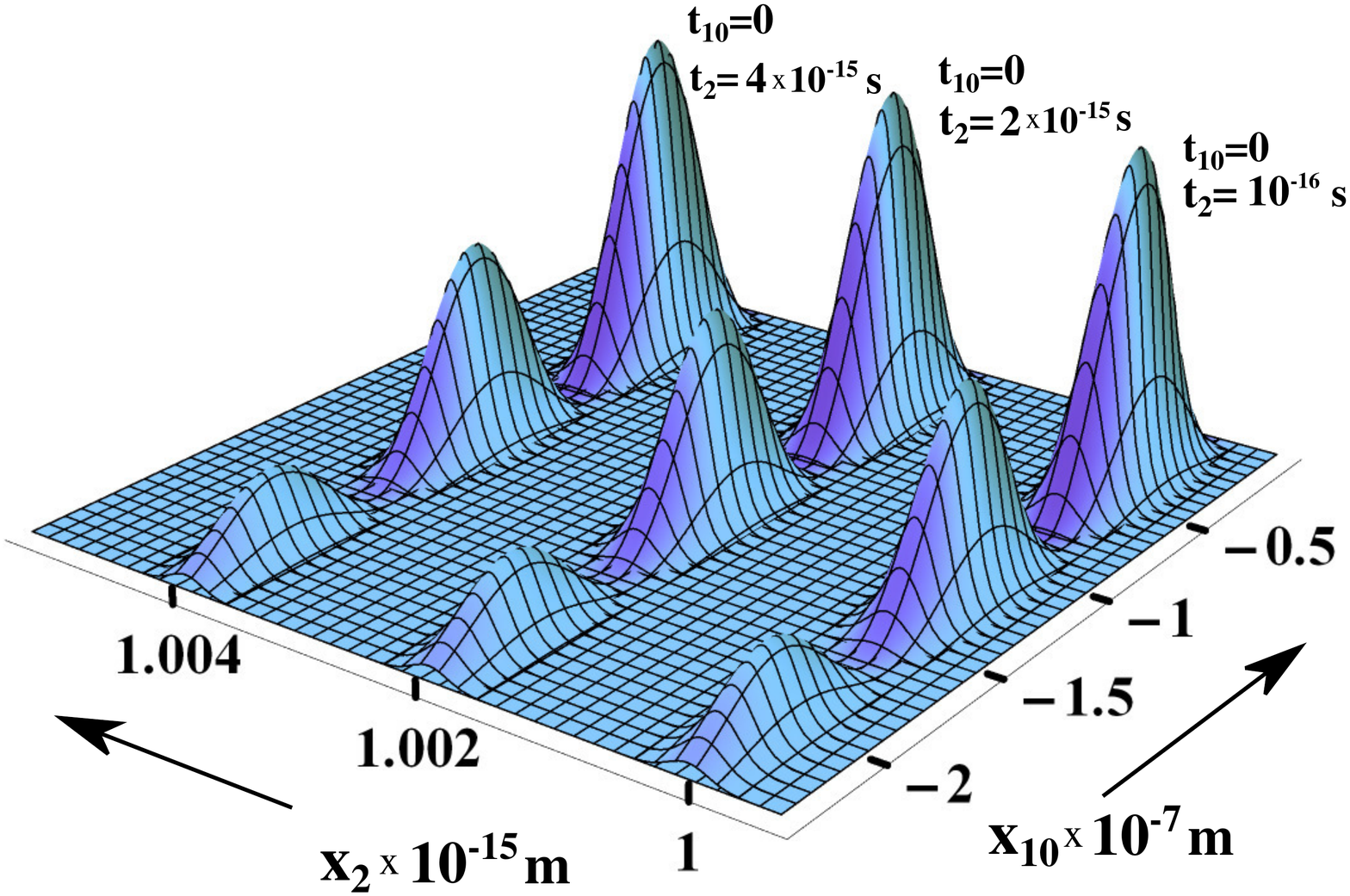}
\caption{One-body mirror PDF snapshots predictions for {\em non-simultaneous} measurement of a rubidium atom reflecting from a mirror of mass $10^{-8}$ kg. The longitudinal coherence length of the atom and mirror are determine by thermal equilibrium at  $T=10^{-7}$ K (released from a Bose-Einstein condensate) and $T=1$ K, respectively. Each snapshot progressively increases the time $t_{2}$.}
\label{fig:largemass}
\end{figure}
\end{center}

One might expect that the small coherence length associated with a mesoscopic/macroscopic mirror mass would not allow the interference shown in fig. \ref{fig:spreadvelocity} (a). This is modeled in fig. \ref{fig:largemass} where snapshots of two-body PDFs are shown at three times for a rubidium atom with $m=1.4 \times 10^{-25}$ kg reflecting from a mirror with $M=10^{-8}$ kg. The spread in wavevectors is determined by thermal equilibrium using $T=10^{-7}$ K for the atom (which is released from a Bose-Einstein condensate) and $T=1$ K for the mirror. The snapshots are for $t_{10}=0$ with $t_{2}= 10^{-16},~2 \times 10^{-15},$ and $ 4 \times 10^{-15}$ s.

This mirror longitudinal coherence length, determined by thermal equilibrium at $T=1$ K, corresponds to the interference in reflection shown in fig. \ref{fig:spreadvelocity} (c), for simultaneous measurement. For non-simultaneous measurements, Fig. \ref{fig:largemass} shows only snapshots over a small range of times $t_{2}$, when the mirror is measured, while the particle has been measured at $t_{10}=0$.

However, varying $t_{2}$ over a much larger range, for a mirror initially at rest, does not change the character of the joint PDF shown in this figure. Initial mirror and particle velocities of $V=1/100$ and $v=3/100$ m/s are used in fig. \ref{fig:largemass}, which results in a beat frequency of $3 \times 10^{5}$ Hz and changes the joint PDF in a manner similar to that shown in fig. \ref{fig:doppler2} along the $x_{2}$ axis. Although these two interfering mirror states (which either reflect or do not reflect the particle) have different speeds, they do not completely separate, as do those shown in the leftmost plot of fig. \ref{fig:regimeB}. This is a consequence of the small difference in mirror speeds due to the large $M/m$ ratio, resulting in a mirror offset which is small compared with the size of the wavegroup.

These results illustrate that a measurement of the particle at one of the minima of the interference pattern, in the joint PDF of fig. \ref{fig:largemass}, yields no possibility of measuring the cm of the mirror for any value of $x_{2}$ and subsequent time $t_{2}$ (within a duration limited by the beat frequency). 

This is to be contrasted with the result shown in fig. \ref{fig:interference}, where simultaneous measurement of the particle and mirror at an interference minimum {\em is} constrained temporally and spatially as follows. First, the duration of the simultaneous interference is finite. For a static mirror, it is essentially determined by the coherence length of the particle divided by its speed, $\approx l_{c}^{particle}/v$ when $l_{c}^{particle} >> l_{c}^{mirror}$. Second, the spatial extent of the interference along the $x_{2}$ axis is not infinite but severely limited by $l_{c}^{mirror}$. Experimental confirmation of the interference in fig. \ref{fig:interference} therefore requires both a small spatial and a short temporal resolving power of the measuring instrument.

Destructive interference, on the other hand, is not similarly constrained. The interference minimum in fig. \ref{fig:largemass} occurs over a range of $x_{2}$ much larger than the mirror coherence length. It is also robust over a wide range of mirror masses.

\subsubsection{Regime B: Doppler effect from a measurement of the particle but not the mirror}
\label{subsec:trace}

Correlated measurements are more difficult to perform than only a measurement on the particle, while making none on the mirror. Predictions of such effects are given by an average of the PDF (or trace) over the mirror coordinate, converting it into a marginal PDF, or for this two-body system a `one-body' PDF \cite{Gottfried}.

In the context of this work, the probability for detecting the particle without measuring the mirror for incident and reflected energy eigenstates states is then manifest as an integral of eqn. \ref{eq:EntangledInterference} over the mirror coordinate (or given by the reduced density matrix obtained as a trace over the mirror coordinates \cite{Tommasini}). Such a procedure, applied to  eqn. \ref{eq:EntangledInterference}, ``washes out'' the one-body interference. Gottfried carries out a similar averaging on the two-body PDF for simultaneous measurement in a momentum-conserving decay after each particle traverses separate double slits with a similar result \cite{Gottfried}.

However, a one-body standing wave interference pattern is certainly expected for the particle reflecting from a static mirror. For a moving mirror, the Doppler shift introduces a time dependence in this pattern. Therefore, the prediction that there is no such interference between incident and reflected particle states, when the mirror is not measured, seems surprising. 

This prediction, however, must be re-evaluated for wavegroups. Consider the averaging procedure when applied to the four PDFs shown in fig. \ref{fig:spreadvelocity}. It is apparent that the one-body interference `washes out' in graph \ref{fig:spreadvelocity}(a) since many cycles in the mirror coordinate of the joint PDF are averaged over, while that is not the case for fig. \ref{fig:spreadvelocity}(c). As shown in fig. \ref{fig:largemass}, the mesoscopic/macroscopic mirror has a coherence length smaller than the fringe spacing determined by an ultracold atom reflecting from it and therefore most closely resembles fig. \ref{fig:spreadvelocity}(c). It is also apparent from this figure that averaging over $x_{2}$ does {\em not} `wash out' the one-body interference. However, this averaging of fig. \ref{fig:spreadvelocity}(c) is only for simultaneous measurement and does not change with time even if the mirror moves (see subsection \ref{sec:asynchDoppler} for more details).

\begin{center}
\begin{figure}
\includegraphics[scale=0.3]{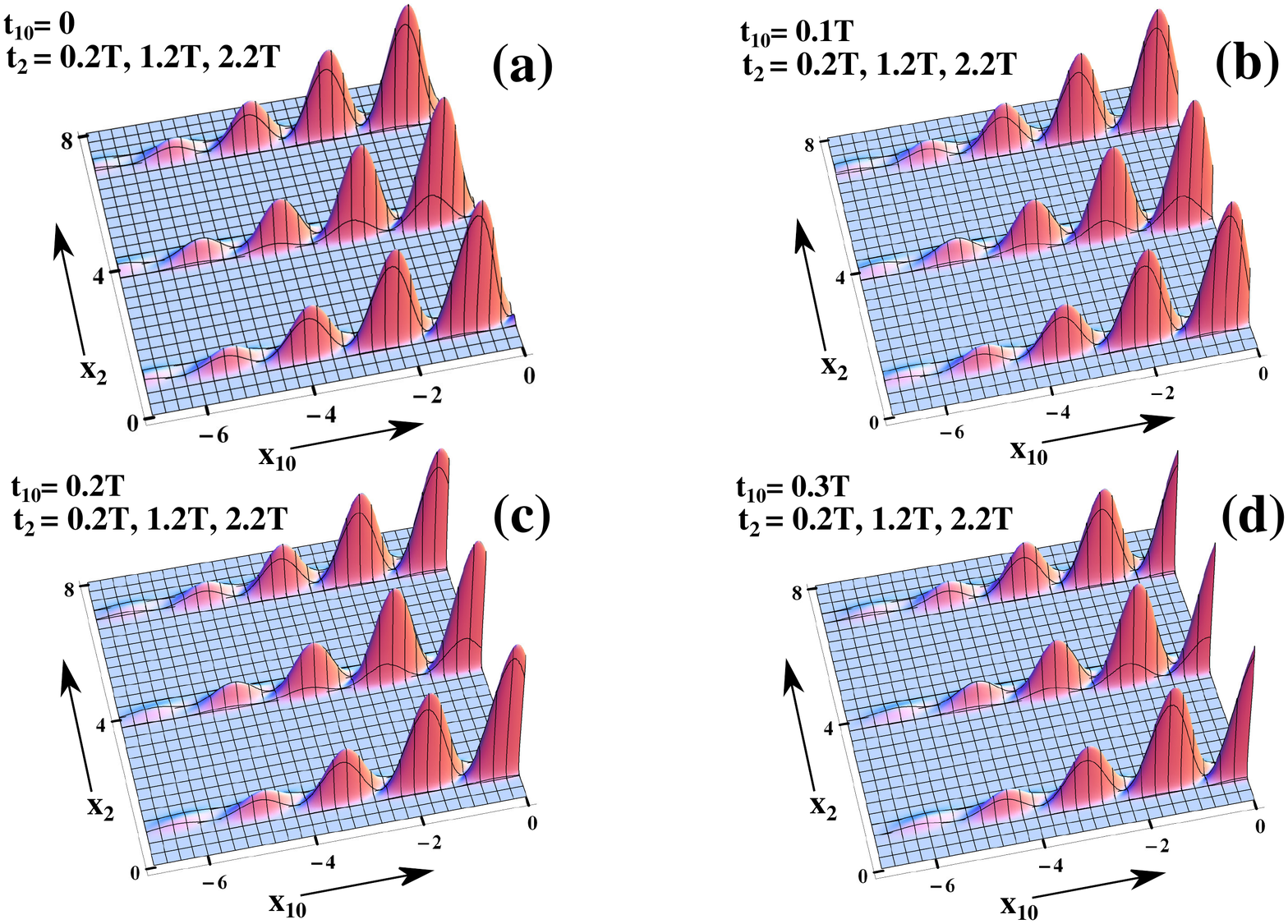}
\caption{Four sequential one-body mirror PDF snapshots of a particle reflecting from a mirror, (a)$\rightarrow$(d), with different $t_{10}$ values while within each plot $t_{2}$ is varied. This illustrates how the result of averaging the PDF over $x_{2}$ is independent of $t_{2}$, leading to the expected one-body interference (derived from the two-body joint PDF) when the mirror is {\em not} measured.}
\label{fig:doppler2}
\end{figure}
\end{center}

For an asynchronous measurement of the particle at $x_{10}$ and $t_{10}$, the joint PDF must be integrated over $x_{2}$ to obtain the marginal PDF for only the particle, if the mirror is not measured. Yet at what time $t_{2}$ must this integration occur for asynchronous measurements? This is addressed in fig. \ref{fig:doppler2}, where four sequential plots are shown for different $t_{10}$ values, while within each plot snapshots at different times $t_{2}$ are shown. The bandwidth parameters used in fig. \ref{fig:doppler2} have been adjusted to model those of a mesoscopic/macroscopic mirror mass. It is apparent from this figure that averaging over $x_{2}$ does not `wash out' the one-body interference for any value of $t_{2}$. That is, the result of averaging over $x_{2}$ is independent of $t_{2}$. Yet variation of the one-body PDF with $t_{10}$ at fixed $x_{10}$ (the Doppler shift) is evident in fig. \ref{fig:doppler2}. As mentioned above, the initial mirror and particle velocities of $V=1/100$ and $v=3/100$ m/s that are used in fig. \ref{fig:largemass} result in a beat frequency for the particle of $3 \times 10^{5}$ Hz.

This effect of the Doppler shift on the particle PDF (shown along the $x_{1}$ axis) differs from that in fig. \ref{fig:doppler}, which illustrates the effect of the Doppler shift in the mirror's PDF (shown along the $x_{2}$ axis). There is no measurable Doppler interference effect for the mirror in  fig. \ref{fig:doppler2}, due to its coherence length being much shorter than the fringe spacing.

The choice of short coherence length mesoscopic/macroscopic mirror and a long coherence length particle substates yields the expected interference for a particle reflecting from a mirror when the mirror is not measured, but requires the use of the two-time formalism (again see subsection \ref{sec:asynchDoppler} for more details). However, such a prediction is limited to a special case of a more general result. For example, this averaging procedure applied to fig. \ref{fig:spreadvelocity}(d), for simultaneous measurement, does not yield the expected particle interference in reflection when the mirror is not measured, since there is no interference over a large range of $x_{1}$ values where it exists in fig. \ref{fig:spreadvelocity}(c).

Consider the opposite case of measurement of the mirror without a measurement of the particle, for the assumptions used in fig. \ref{fig:doppler2}. In this case, the averaging is over $x_{1}$, which washes out the fringes. Therefore, measurement of just the mirror results in no interference. This interference for the particle, but not the mirror, is that expected for a classical wave reflecting from a mirror. Yet, it is a special case rather than a general result since it depends on the coherence length assumptions.

\section{Discussion}
\label{sec:discussion}

\subsection{Prolonging interference with a perceptible fringe spacing for a mesoscopic/macroscopic mirror}
\label{sec:prolong}

Synchronous correlated interference in reflection is temporally and spatially limited by the particle-mirror coherence lengths as the incident and reflected two-body wavegroups separate. We have presented a simple model illustrating the magnitude of these effects. 

Measurement of the particle first can prolong interference of the mirror, which is measured later. The resulting two one-body mirror states travel in the same direction but with momenta and energies differing by that either exchanged or not exchanged with the particle. This superposition state can either interfere or not depending on the coherence length of the mirror and the time between measurements. 

For an asynchronous measurement with a small mirror coherence length, interference persists only for a small momentum exchange. Since these states have different energies their interference results in a `beat' or time dependent fringe pattern for the mirror PDF. For longer times between measurements, these two mirror wavegroups then separate beyond their coherence lengths, due to their different average momenta. This then results in a spatially separated superposition state without interference. 

A microscopic particle reflecting from a mesoscopic/macroscopic mirror at rest is well suited to prolong this one-particle interference of the two mirror states for two main reasons. First the two mirror wavegroups maintain overlap longer since their difference in speeds is $\Delta V \approx 4mv/M$, while their coherence length is $l_{c}^{thermal}$. This results in an overlap time $\approx h \sqrt{M}/(4mv\sqrt{2k_{B}T})$. Second their energy difference is negligible for a stationary mirror. That is, the ``beat'' frequency goes to zero in the limit of large mirror mass while the difference in momenta of these mirror states remains small but finite. 

It is this difference in mirror momenta which leads to the phase difference $\Delta K~x_{2}$ when superposing these two mirror wavefunctions. That is, the mirror fringe spacing is only determined by the change in particle momentum since conservation of momentum in reflection requires $\Delta K=-\Delta k$. The fringe structure of the mesoscopic/macroscopic mirror is then commensurate with that of the microscopic particle.

Double slit interference with a massive particle, on the other hand, superposes two one-body states with the same momentum whose difference in phase, $K \Delta x$, is due to the difference in path lengths $\Delta x$ from each slit to the measurement point times an extremely large wavevector (rather than the small $\Delta K$ for the mirror reflecting a microscopic particle). This results in an imperceptible fringe spacing on the mesoscopic/macroscopic object traversing the double slits.

The interference of the mesoscopic/macroscopic mirror states does not require utilizing standard division of amplitude nor division of wavefront interferometric methods. It does, however, use measurement acting as a beamsplitter to generate two mirror states which interfere for times greater than that for synchronous measurement. In addition, path lengths need not be carefully matched for interference to be manifest. These are distinct experimental advantages over division of amplitude or wavefront interferometers, particularly for a mesoscopic/macroscopic mirror mass.

\subsection{Measurement theory}
\label{sec:measurementtheory}

Asynchronous measurement contains some subtle issues. Foremost among them is what constitutes a measurement of the particle, after which no interaction with the mirror occurs. An experimental realization of such a measurement might consist of the particle-mirror reflection occurring only along the x-axis while another particle called the ``probe'' moves along the y-axis, as illustrated in fig. \ref{fig:overview}. A measurement then involves the probe absorbing, being absorbed by, being scattered by, or scattering from the particle. The effect of this measurement on the particle (e.g. back action) has no effect on the mirror substate by the no interaction assumption. 

The operational definition of such a measurement, when the particle is in the correlated interference region, is then that which causes splitting of the mirror state. This could result in the observation of prolonged interference of the mirror states or observation of these states when they no longer overlap (this is discussed in more detail in section \ref{sec:measuring}) . Note that the probe need not interact with the particle at any particular location apart from it occurring within this correlated interference region (and not at a PDF minimum). 

Related issues involve modeling this sequential measurement. In the Copenhagen interpretation, a measurement of the particle's position collapses its substate into an eigenstate of that position operator. The assumption of no interaction after the first measurement then requires that the particle does not reflect a second time from the mirror's substate. The state of the particle after measurement is therefore irrelevant to the subsequent evolution of the mirror substate and is treated as such in the formalism presented above by fixing the particle coordinates in the two-body function at its measurement position and time. A more realistic model would account for a measurement which occurs over a distance $\Delta x_{1}$ rather than at a point in space. Such a treatment is beyond the scope of this work.

This assumption of the lack of interaction can be relaxed to further study the implications of the collapse postulate. To illustrate this let $\Psi_{1}[x_{1},t_{1},x_{2},t_{2}]$ be the solution to the Schr\"odinger equation for the incident uncorrelated plus the reflected correlated wavegroups. Also let the first measurement be of the particle in the correlated interference regime at $x_{1}=x_{10}$ and time $t_{10}$. The particle substate then collapses into an eigenstate of the position operator and therefore generates the two-body uncorrelated wavefunction $\Psi_{20}=\Psi[x_{1},t_{10}] \Psi_{1}[x_{10},t_{10},x_{2},t_{10}]$ where $\Psi[x_{1},t_{10}]=\delta[x_{1}-x_{10}]$. To determine the effects of allowing interaction (a second reflection) before a later measurement of the mirror, the solution to the Schr\"odinger equation $\Psi_{2}[x_{1},t_{1},x_{2},t_{2}]$ for $t_{1}>t_{10}$ and $t_{2}>t_{10}$ is needed. This is determined from the initial condition $\Psi_{20}$ along with the boundary condition at the mirror surface. However, the particle must not be absorbed in the measurement process since it is now allowed to interact again with the mirror after its initial measurement.

As a particular example, consider measuring the particle at a position in the correlated interference region given by the middle snapshot in fig. \ref{fig:interference}. The particle-mirror state just after this measurement is a cut along the $x_{2}$ axis at $x_{1}=x_{10}$ of the particle-mirror wavefunction associated with this PDF, due to the delta function nature of this collapse. This wavefunction in the $(x_{1},x_{2})$ plane is then the initial condition used in solving the two-body Schr\"odinger equation for the subsequent particle-mirror state. The reflection boundary condition at $x_{1}=x_{2}$ must also be satisfied. The narrow particle substate along the $x_{1}$ axis will lead to its ``diffusion'' and subsequent ``re-reflection.''

Note that if the particle is measured a second time, before a measurement of the mirror, then the procedure just mentioned must be repeated. Measurement-induced multiple reflections then occur and are, in themselves, of fundamental interest.

\subsection{Asynchronous measurement: prediction of the Doppler effect}
\label{sec:asynchDoppler}

A time dependent interference pattern is expected classically when a harmonic wave reflects from a moving mirror. For a stationary mirror the incident and reflected waves form a standing wave which is measured by placing a detector at a fixed position where the incident and reflected waves overlap \cite{Wiener}.  Motion of the mirror introduces a Doppler shift in the reflected wave which generates time dependence in this interference. The detector fixed in space then measures a time dependent intensity variation.

Consider the analogous process for the particle-mirror system. The incident and reflected two-body wavefunctions have the same total energy in reflection from a moving mirror. Interference in the synchronous formalism involves superposing these two wavefunctions, each with the same frequency. The result, given by eqn. \ref{eq:EntangledInterference} with $t_{1}=t_{2}$, is then independent of time. 

To illustrate this, consider the middle snapshot of fig. \ref{fig:interference}. For different snapshots (not shown), all of which still partially overlap spatially with this middle snapshot, the envelope of the PDF waveform moves through the $(x_{1},x_{2})$ plane, yet the maxima and minima of the fringes for these different snapshots remains in the same location. 

The Doppler shift for the particle-mirror system, analogous to a classical wave reflecting from a moving mirror, involves only measuring the particle flux at a fixed position with no measurement of the mirror. Not measuring the mirror involves averaging the two-body PDF (trace) over the mirror coordinates. The mirror coherence length determines if this averaging ``washes-out'' the interference. 

An example, for the synchronous case is shown in fig. \ref{fig:spreadvelocity} (a). However, a smaller mirror coherence length results in a PDF which, when averaged over $x_{2}$, maintains the interference along $x_{1}$, as is illustrated in fig. \ref{fig:spreadvelocity} (c). The PDF for simultaneous correlation, with a coherence length associated with that shown in fig. \ref{fig:spreadvelocity} (c), then generates only static interference and not the expected time dependent interference effect associated with the Doppler shift.

However, as shown in subsections \ref{subsec:beat} and \ref{subsec:trace}, the two-time formalism predicts a Doppler interference effect similar to that classically when measuring only the particle, but only as a special case (for certain coherence lengths) of a more general result. This is a consequence of using the two-time formalism to parse the total energy and therefore the frequency of the wavefunction into terms associated with the particle and mirror rather than only using the expression for the total energy.

\subsection{Decoherence}
\label{sec:decoher}

Interference in a typical one-body interferometer `washes-out' as the fringe spacing becomes imperceptible with increasing mass. Another mechanism that eliminates interference is a measurement to determine along which path the object traveled through the interferometer. The former issue is mitigated using two-body reflection of a microscopic particle from a mesoscopic/macroscopic object, as described above, while the latter is the first topic of this section.

\subsubsection{Path information}
\label{sec:path}

A method to determine path information and therefore destroy interference utilizes a `Heisenberg microscope,' where an additional particle scatters from the object traversing the interferometer. If the wavelength of this particle, $\lambda$, is smaller than the spatial separation along the two paths, $\Delta x$, then path information can be obtained in principle from the state of this scattered probe particle and interference is destroyed. This was verified using a three grating Mach-Zehnder interferometer traversed by an atom and using a photon as the probe to determine path information \cite{chapman}. The contrast of the interference, as a function of $\Delta x$, was shown to drop dramatically for $\Delta x>\lambda/2$.

Generating such decoherence for particle-mirror interference involves using a probe particle to determine the position of the mirror during interference (after measurement of the particle). There are two distances that the mirror could have moved: that associated with and that without reflection. If the probe particle can resolve the difference in these distances then the `path' associated with reflection is distinguishable from that without reflection. 

For a synchronous measurement of a near stationary mirror with $m/M<<1$ and $V/v<<1$, the mean difference in these mirror distances is $\Delta x\approx 2l_{c}^{particle}m/M$, which when equal to the wavelength of a probe particle of mass $m^{*}$ then requires that the probe have velocity $v_{probe}\approx hM/(4l_{c}^{particle}mm^{*})$. For asynchronous measurement $\Delta x\approx 2v\Delta tm/M$ requiring a probe velocity $v_{probe}\approx hM/(2mm^{*}v\Delta t)$, where $\Delta t=t_{2}-t_{1}$.

This calculation can also be used to approximate the effect of such decoherence on the mirror from the collisions with gas atoms forming the thermal bath which surrounds the mirror if these gas atoms are considered to be the probe which determines path information and they scatter from only one of the superposed mirror states. The interference in reflection described above is then maintained if there is a small probability of these atoms having the velocity $v_{probe}$. For a mesoscopic mirror, $M\approx 10^{-10}$ kg, and both microscopic particle and probe atoms of mass $\approx 10^{-25}$ kg, $v_{probe} \approx 10^{6}$ m/s. Atoms with such thermal speeds are not likely to be found in a low temperature gas.

This decoherence mechanism can also be applied to the emission of three examples are used by the mirror during interference. If the photon's wavelength is small enough to allow for the location of the slab associated with one of these positions, interference is destroyed since it can then be determined at which position the reflection occurred \cite{cronin}. For an ultra cold atom reflecting from a mirror of mass $M \approx 10^{-10}$ kg, $\Delta x\approx 2v\Delta tm/M$, requiring a wavelength of $\approx 10^{-15}$ m for $\Delta t=1$ s. The probability of such a thermal photon emission is small.

However, for such a probe to destroy mirror interference it must interact with only one of the mirror states (how this could be accomplished with closely spaced mirror states is not obvious). If not, then the probe (with a coherence length longer than the separation of the mirror states) itself is placed into a superposition state of having interacted with both mesoscopic/macroscopic mirror states. This results in probe interference, which can be used as a method to determine that the mirror is in a superposition state, as discussed in section \ref{sec:measuring},

These calculations indicate that decoherence associated with obtaining path information can be mitigated with a judicious choice of parameters. This is due in part to the displacements associated with the interfering mirror substates being proportional to $m/M$. To measure such small displacements requires probe particles with proportionally smaller wavelengths which are typically not found in the environment.

\subsubsection{Environmental decoherence}
\label{sec:environ}

Another decoherence mechanism is through coupling with the many degrees of freedom of the surrounding environment. Zurek \cite{zurek} predicted exponential decay of the off-diagonal density matrix element terms for a mesoscopic/macroscopic object in a superposition state, $\exp[-t/t_{D}]$, with time constant $t_{D}=t_{R}(\lambda_{T}/\Delta x)^{2}$, where $t_{R}$ is the thermal relaxation time, $\Delta x$ is the difference in distance between these two states, and $\lambda_{T}$ is the thermal de Broglie wavelength of the mesoscopic/macroscopic object. The decay is into a mixed rather than a pure state.

Applying this decoherence model to the mesoscopic/macroscopic mirror states after measurement of the particle, as described above, yields $t_{D}/t_{R}\approx M h^{2}/(8k_{B}T(mv \Delta t)^{2})$. Reflection of an atom from a mirror of mass $M\approx 10^{-8}$ kg at $T=1$ K yields $t_{D}\approx t_{R} 10^{5}$ for $\Delta t=1$ s. However, $t_{R}$ can vary dramatically depending on environmental and material properties along with the dimensional values of the mirror. Macroscopic yet small masses can have short thermal relaxation times. Nevertheless, it appears that environmental decoherence of the mesoscopic/macroscopic mirror states generated from asynchronous measurement can be made either negligible or non-negligible in this model of decoherence. 

This decoherence time was derived to show that mesoscopic/macroscopic bodies in superposition states quickly decohere. Therefore, any interference due to such superposition should not be observed. On the other hand, asynchronous measurement generates a superposition of macroscopic mirror states which mitigate this environmental decoherence, due to their small spatial separation.

\subsubsection{Decoherence overview}
\label{sec:overview2}

It has been shown that correlated interference for a microscopic particle reflecting from a mesoscopic/macroscopic mirror, does not disappear with increasing mirror mass, is difficult to eliminate via a path measurement, and can be made insensitive to environmental decoherence. It should also be pointed out that the robust character of interference for objects with many degrees of freedom is reinforced by measurements which demonstrate that even if the size of the object is larger than both the coherence length and deBroglie wavelength, interference can still be observed \cite{cronin}.

\subsection{Measuring the mirror in a superposition state}
\label{sec:measuring}

A third particle, reflecting from the mirror after asynchronous measurement, can act as a probe to confirm that the mirror is in a superposition state. Such reflection is again described by a two-body wavefunction, if the measurement of the initial particle has eliminated this particle from interacting with both the split mirror states and the probe. The result is then a sum of two-body solutions to the Schr\"odinger equation, one for the probe reflecting from each mirror state. Wavegroups are then formed from these harmonic states as described above.   

The interference in reflection of these two-body wavegroups is robust since overlap is maintained as they both travel in the same direction. Measuring the probe, without any measurement of the mirror, does not destroy the probe interference for sufficiently small mirror coherence lengths, as described in subsection \ref{sec:asynchDoppler}. Therefore, the more difficult correlation measurement of both the probe and mirror need not be performed.

The coherence length of the probe need only be larger than the very small displacement of the mirror states for this probe interference to be exhibited, while the phase difference between these probe states is determined by the mass of the probe. With such interference, the probe cannot distinguish between the mirror states as discussed in section \ref{sec:path}.  Verification of interference from only the reflected probe is then a method to measure the mesoscopic/macroscopic mirror in a superposition state, without any `direct' measurement of the mirror. Details of such a calculation are beyond the scope of this foundational work.

\section{Summary}
\label{sec:summary}

Asynchronous measurement in quantum mechanics has been treated for a particle reflecting from a mirror. The results presented assume measurement of the particle first, collapse of the particle substate, and a later measurement of the mirror while no particle-mirror interaction occurs between the times of these measurements. The two-body Schr\"odinger equation, solved with standard techniques, is parsed to completely separate the variables associated with the particle from those of the mirror, so that when the particle is measured only its parameters are fixed while those of the mirror can continue to evolve in time until the mirror is measured. 

The subsequent interference of the mirror states which have and have not reflected the particle is in many ways familiar from examples in quantum mechanics in which an outcome can be achieved in indistinguishable ways. Yet this process of splitting differs by using measurement as its mechanism. This is not possible in a one-body system. In addition, the example used here of a microscopic particle reflecting from a mesoscopic or macroscopic mirror illustrates why these interference effects do not vanish with increasing mirror mass.

{\em Without} such measurement, the two-body wavegroup time evolves into one in which the wavegroup has completely reflected and then no longer overlaps with the incident wavegroup. Correlated interference then disappears since there is then no uncertainty about the particle and mirror belonging to the incident or reflected states. 

Correlated interference in reflection is an example of one of the simplest interferometers, utilizing neither division of amplitude nor division of wavefront methods to generate interference. In addition, path lengths need not be carefully matched for interference to be manifest.

The complexity of {\em asynchronous} correlated interferometry is revealed as the coherence lengths of the particle and mirror substates are varied. Examples are: (1) When both are small then essentially classical reflection results. Measurement of the particle either occurs after {\em or} before reflection, resulting in a later measurement of the mirror either having recoiled {\em or} not recoiled.  (2) When both are large, compared with the particle's wavelength, then measurement of the particle in the correlated interference region acts as a beamsplitter generating mirror states which have {\em and} have not reflected the particle. (3) Measurement of the particle only, for short mirror and long particle coherence lengths, yields the expected interference due to the Doppler effect but only as a special case of a more general result.

Other issues of asynchronous measurements presented include: transfer of coherence between the particle and mirror, the effects of substate coherence on asynchronous interference, distortion of the two-body wavegroups due to reflection, interference effects on mesoscopic/macroscopic mirrors, and the prolonging of interference of the mirror substate when the particle is measured first in the region of correlated interference. Finally, the superposition of mesoscopic/macroscopic mirror states was shown to be insensitive to environmental decoherence.

Some consequences of asynchronous measurements not presented include: multiple sequential particle measurements before a measurement of the mirror (two-body quantum Zeno effect), treatment of asynchronous measurement in other interpretations of quantum mechanics, two-body effects for spatially varying potentials, reflection of a massless particle, dealing with an inelastic collision, and both synchronous and asynchronous two-body treatment of quantum tunneling. Perhaps the one modification which would result in a more familiar interferometer is to incorporate a partially transmitting, rather than totally reflecting, mirror. One example is of a thin aluminum slab whose two surfaces partially retro-reflect a neutron. This then leads to a two-body interferometer analogous to the one-body thin-film interference of light. Finally, the more likely domain in which these prediction will be tested is in the reflection of a microscopic particle from a microscopic ``mirror,'' the details of which have also not been discussed.

There is little experimental support for asynchronous correlation interferometry. Yet, it involves perhaps the most elementary of only a few exact solutions to the two-body Schr\"odinger equation. It is surprising that such theoretical simplicity is not reflected experimentally.

Measurement in quantum mechanics plays a role different from that in any other physical theory and its interpretation remains controversial. It is shown here that measurement can act as a beamsplitter, which, under the appropriate conditions, can prolong the transient interference associated with reflection in a two-body quantum system. Although far from being comprehensive, these results also indicate a direction, heretofore unexplored, for further research in understanding asynchronous quantum correlation, probing decoherence, and extending quantum measurements to larger masses.

\begin{acknowledgments}
The authors would like to thank Professors C. Durfee and J. Scales for useful comments, along with referee guidance.
\end{acknowledgments}

\section{Appendix}
\label{sec:appendix}

Without collapse the two-body wavefunction, expressed in terms of the two times, evolves and conserves probability. The probability of measuring the particle at $(x_{1},t_{1})$ and the mirror at $(x_{2},t_{2})$ is given by $\iint PDF[x_{1},t_{1},x_{2},t_{2}] dx_{1} dx_{2}$ with the joint PDF determined by the solution of equation \ref{eq:Schreqn} as $\Psi \Psi^{*}$ but expressed in terms of the two time variables in the particle-mirror system. Using this equation, conservation of probability can then be expressed locally as,
\begin{eqnarray}
\partial_{t_{1}}PDF[x_{1},t_{1},x_{2},t_{2}] +\partial_{t_{2}} PDF[x_{1},t_{1},x_{2},t_{2}] +\notag \\
\partial_{x_{1}} j_{1}[x_{1},t_{1},x_{2},t_{2}] +\partial_{x_{2}} j_{2}[x_{1},t_{1},x_{2},t_{2}]=0,
\label{eq:consProb}
\end{eqnarray}
where $j_{1}[x_{1},t_{1},x_{2},t_{2}]=\hbar (\Psi^{*} \partial_{x_{1}} \Psi-\Psi \partial_{x_{1}} \Psi^{*})/(2 i m)$ and  $j_{2}[x_{1},t_{1},x_{2},t_{2}]=\hbar (\Psi^{*} \partial_{x_{2}} \Psi-\Psi \partial_{x_{2}} \Psi^{*})/(2 i M)$. While the expressions for these current densities appear similar to that for one particle systems there are subtle but important differences for a two body system \cite{currentdensity}.

Multiplying equation \ref{eq:consProb} by $dx_{1} dx_{2}$, integrating over the segment from $a$ to $b$ along the x-axis ($a \leq x_{1} \leq b$ and $a \leq x_{2} \leq b$), and then rearranging terms yields a solution to equation \ref{eq:consProb} if
\begin{eqnarray}
\partial_{t_{1}} \int_{a}^{b} PDF[x_{1},t_{1},x_{2},t_{2}] dx_{1} +\notag \\ j_{1}[b,t_{1},x_{2},t_{2}]-j_{1}[a,t_{1},x_{2},t_{2}]=0 \notag
\label{eq:consProb2a}
\end{eqnarray}
and
\begin{eqnarray}
\partial_{t_{2}} \int_{a}^{b} PDF[x_{1},t_{1},x_{2},t_{2}] dx_{2} +\notag \\ j_{2}[x_{1},t_{1},b,t_{2}]-j_{2}[x_{1},t_{1},a,t_{2}]=0  \notag.
\label{eq:consProb2b}
\end{eqnarray}
These equations indicate that the time rate of change of probability within the $\overline{ab}$ segment on the x-axis is determined separately by a net change in particle and mirror probability fluxes in that region, which is similar to conservation of probability in a one-body system. Now, however, probability of the two-body system is conserved for the mirror even when $x_{1}$ and $t_{1}$ are fixed. The asynchronous procedure described here therefore maintains conservation of probability for the mirror after the particle has been measured.


\begin{references}
\bibitem{beat} A beat frequency is usually associated with an AC coupled voltage from a detector placed at the output port of an interferometer which combines waves of different frequencies. This voltage, proportional to the intensity, has the same trigonometric dependence as the joint PDF given in eqn. \ref{eq:EntangledInterference}. However since $\sin^{2}\Omega t=(1-\cos2 \Omega t)/2$, the frequency of this AC coupled output voltage is $2\Omega$.
\bibitem{currentdensity} In the particle-mirror system the incident and reflected wavevectors for the particle (or mirror) are not necessarily equal and opposite as they are in the cm-rel system. The probability current density to the left of the mirror then exhibits spatial and temporal interference due to the effect this difference in wavevectors has on the incident and reflected wavefunctions. This results in a non-additive current density, a consequence of the increasing space being generated as the mirror moves away from the particle. The net flux  must account for the build-up of probability within this ``new'' space.
\end{references}
\end{document}